\documentclass[preprint,texmf,texlive=2011,UKenglish,txfonts,siunitx=false]{atlasdoc} 
\pdfoutput=1
\usepackage{atlasphysics}
\usepackage[biblatex=false,subfigure=true]{atlaspackage}
\usepackage{multirow}
\usepackage{cite}
\usepackage{bm}

\newlength{\figwidth}
\DeclareGraphicsExtensions{.pdf}

\AtlasTitle{Constraints on the off-shell Higgs boson signal strength in the high-mass $ZZ$ and $WW$ final states with the ATLAS detector}

\author{The ATLAS Collaboration}

\PreprintIdNumber{CERN-PH-EP-2015-026}
\AtlasJournal{Eur. Phys. J. C} 
\date{\today}

\newcommand{\KHmZZ}{\ensuremath{\text{K}^{H^*}(m_{ZZ})}}
\newcommand{\KHggmZZ}{\ensuremath{\text{K}_{gg}^{H^*}(m_{ZZ})}}

\newcommand{\KHmVV}{\ensuremath{\text{K}^{H^*}(m_{VV})}}
\newcommand{\KHggmVV}{\ensuremath{\text{K}_{gg}^{H^*}(m_{VV})}}
\newcommand{\KBmVV}{\ensuremath{\text{K}^\text{B}(m_{VV})}}
\newcommand{\Kratio}{\ensuremath{\text{R}^B_{H^*}}}
\newcommand{\Widthratio}{\ensuremath{\Gamma_H/\Gamma_H^\text{SM}}}
\newcommand{\muonshell}{\ensuremath{\mu_\text{on-shell}}}
\newcommand{\muoffshell}{\ensuremath{\mu_\text{off-shell}}}

\newcommand{\CLs}{\ensuremath{CL_s}}

\newcommand{\ww}{WW \to e\nu\,\mu\nu}
\newcommand{\zzn}{ZZ \to 2\ell\,2\nu}

\newcommand{\hwwonshell}{\ensuremath{0.04\pm0.03}\ } % onshell contamination
\newcommand{\hwwWWcrpur}{\ensuremath{46\%}\ }   % ww cr qqww purity
\newcommand{\hwwWWcrobs}{\ensuremath{8007}\ } % ww cr obs data
\newcommand{\hwwWWcrt}{\ensuremath{39\%}\ } % ww cr top contam
\newcommand{\hwwWWcrNF}{\ensuremath{1.03\pm0.11}} % ww nf
\newcommand{\hwwTcrpur}{\ensuremath{96\%}\ } % top cr top purity
\newcommand{\hwwTcrobs}{\ensuremath{13498}\ } %top cr obs data
\newcommand{\hwwTopcrNF}{\ensuremath{1.03\pm0.04}} % ww nf
\newcommand{\hwwstfrac}{\ensuremath{22\%\ }} % st fraction of top
\newcommand{\hwwstsys}{\ensuremath{20\%\ }} % st norm sys
\AtlasAbstract{%
Measurements of the $ZZ$ and $WW$ final states in the mass range above the $2m_Z$ and $2m_W$ thresholds provide a
unique opportunity to measure the off-shell coupling strength of the Higgs boson.
This paper presents constraints on the off-shell Higgs boson event yields normalised to the Standard Model prediction (signal strength) in 
the $ZZ \to 4\ell$, $ZZ\to2\ell2\nu$ and $WW\to e\nu\mu\nu$ final states.
The result is based on $pp$ collision data collected by the ATLAS experiment at the LHC,
corresponding to an integrated luminosity of $20.3$ fb$^\text{-1}$ at a collision energy of $\sqrt{s}=8$~TeV.
Using the $CL_s$ method, the observed 95\% confidence level (CL) upper limit on the off-shell signal strength is in the range
5.1--8.6, with an expected range of 6.7--11.0. In each case the range is determined by varying the unknown $gg\to ZZ$ and 
$gg\to WW$ background K-factor from higher-order QCD corrections between half and twice the value of the known signal K-factor.
Assuming the relevant Higgs boson couplings are independent of the energy scale of the Higgs production,
a combination with the on-shell measurements yields
an observed (expected) 95\% CL upper limit on $\Gamma_H/\Gamma_H^{\mathrm{SM}}$
in the range 4.5--7.5 (6.5--11.2)  
using the same variations of the background K-factor. 
Assuming that the unknown $gg\rightarrow VV$ background K-factor is equal to the signal K-factor,
this translates into an observed (expected) 95\% CL 
upper limit on the Higgs boson total width of 22.7 (33.0)~MeV.
}

\begin{document}

%\tableofcontents
%-------------------------------------------------------------------------------
\section{Introduction}
\label{sec:introduction}
The observation of a new particle in the search for the Standard Model
(SM) Higgs boson at the LHC, reported by the
ATLAS~\cite{paper2012ichep} and CMS~\cite{CMSpaper} Collaborations,
is a milestone in the quest to understand electroweak symmetry
breaking. Precision measurements of the properties of the new boson are of critical importance. 
Among its key properties are the couplings to each of the SM fermions and bosons,
for which ATLAS and CMS presented results in Refs.~\cite{Aad:2013wqa,Khachatryan:2014jba}, 
and spin/CP properties, studied by ATLAS and CMS in Refs.~\cite{Aad:2013xqa,Khachatryan:2014kca}.

The studies in Refs.~\cite{Kauer:2012hd,Caola:2013yja,Campbell:2013una,Campbell:2013wga} 
have shown that the high-mass off-peak regions beyond $2 m_{V}$ ($V=Z,W$), well above the measured resonance mass of about 125 GeV~\cite{Aad:2014aba,Khachatryan:2014jba}, 
in the $H \to ZZ$ and $H \to WW$  channels 
are sensitive to Higgs boson production through off-shell and background interference effects.
This presents a novel way of characterising the properties of the Higgs boson in terms of the off-shell 
event yields, normalised to the 
SM prediction (referred to as signal strength $\mu$), and the associated off-shell Higgs boson couplings.
Such studies provide sensitivity to new physics that alters the interactions between the Higgs boson 
and other fundamental particles in the high-mass
region~\cite{Englert:2014aca,Cacciapaglia:2014rla,Azatov:2014jga,Ghezzi:2014qpa,Buschmann:2014sia,Gainer:2014hha,Englert:2014ffa}.
This approach was used by the CMS Collaboration~\cite{Khachatryan:2014iha} to set an indirect limit on the Higgs boson total width.
The analysis presented in this paper is complementary to direct searches for Higgs boson to invisible 
decays~\cite{Aad:2014iia,Chatrchyan:2014tja} and to constraints coming from the Higgs boson coupling tests~\cite{Aad:2013wqa,Khachatryan:2014jba}.

This paper presents an analysis of the off-shell signal strength in the $ZZ \to 4\ell$, $ZZ \to 2\ell2\nu$ 
and $WW \to e\nu\,\mu\nu$ final states ($\ell=e,\mu$).
It is structured as follows: Sect.~\ref{sec:theory} discusses the key theoretical considerations
and the simulation of the main signal and background processes.
Sections~\ref{sec:4l}, \ref{sec:2l2n} and \ref{sec:lnln} give details for the analysis in the $ZZ \to 4\ell$, 
$ZZ \to 2\ell2\nu$ and $WW \to e\nu\,\mu\nu$ final states, respectively. 
The dominant systematic uncertainties are discussed in Sect.~\ref{sec:systematics}.
Finally the results of the individual analyses and their combination are 
presented in Sect.~\ref{sec:results}.

The ATLAS detector is described in Ref.~\cite{atlas-det}. The present analysis is
performed on $pp$ collision data corresponding to an integrated luminosity of $20.3$ fb$^\text{-1}$ at a collision energy of $\sqrt{s}=8~\TeV$. 

%-------------------------------------------------------------------------------
\section{Theoretical predictions and simulated samples}
\label{sec:theory}
\begin{figure}[!ht]
\begin{center}
\subfigure[]{\includegraphics[height=0.20\figwidth]{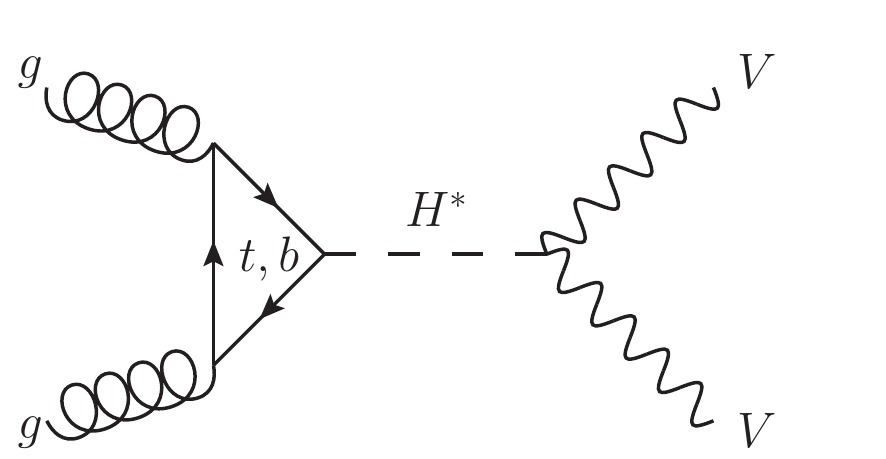}\label{fig:feynman_diagram_ggHVV}}
\hfill
\subfigure[]{\includegraphics[height=0.20\figwidth]{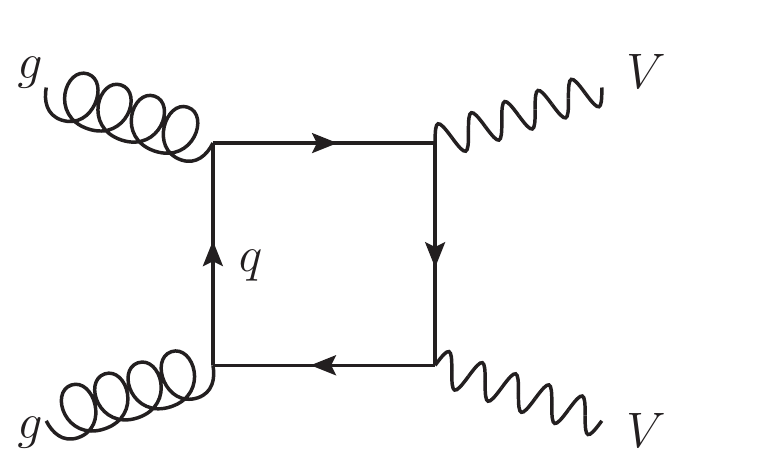}\label{fig:feynman_diagram_ggVV}}
\hfill
\subfigure[]{\includegraphics[height=0.20\figwidth]{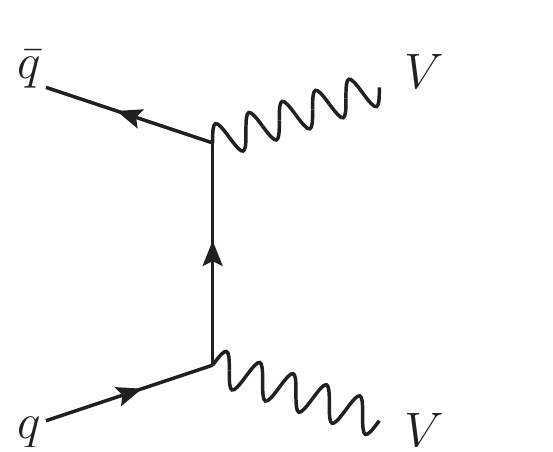}\label{fig:feynman_diagram_qqVV}}
\caption{The leading-order Feynman diagrams for \subref{fig:feynman_diagram_ggHVV} the $gg \to H^* \to VV$ signal,
\subref{fig:feynman_diagram_ggVV} the continuum $gg \to VV$ background and \subref{fig:feynman_diagram_qqVV} the $q \bar{q} \to VV$ background.
}
\label{fig:feynman_diagrams}
\end{center}
\end{figure}

%-------------------------------------------------------------------------------
The cross-section $\sigma_\text{off-shell}^{gg \to H^* \to VV}$ for the off-shell Higgs boson production through gluon fusion with subsequent decay into vector-boson 
pairs,\footnote{%
In the following the notation $gg \to (H^* \to) VV$ is used for the full signal+background process 
for $VV=ZZ$ and $WW$ production, including the Higgs boson signal (S) $gg \to H^* \to VV$ process, 
the continuum background (B) $gg \to VV$ process and their interference.
For vector-boson fusion (VBF) production, the analogous notation VBF $(H^*\to) VV$ is 
used for the full signal plus background process, 
with VBF $H^*\to VV$ representing the Higgs boson signal and VBF $VV$ denoting the background.}
as illustrated by the Feynman diagram in Fig.~\ref{fig:feynman_diagram_ggHVV},
is proportional to the product of the Higgs boson couplings squared for production and decay.
However, unlike the on-shell Higgs boson production, $\sigma_\text{off-shell}^{gg \to H^* \to VV}$ is independent 
of the total Higgs boson decay width $\Gamma_H$~\cite{Kauer:2012hd,Caola:2013yja}. 
Using the framework for  Higgs boson coupling deviations as described in Ref.~\cite{Heinemeyer:2013tqa},
the off-shell signal strength in the high-mass region selected by
the analysis described in this paper at an energy scale $\hat{s}$, $\mu_\text{off-shell}(\hat{s})$, can be expressed as:
\begin{linenomath}
\begin{equation}
  \mu_\text{off-shell}(\hat{s}) \equiv \frac{\sigma_\text{off-shell}^{gg \to H^* \to VV}(\hat{s})}{\sigma_\text{off-shell, SM}^{gg \to H^* \to VV}(\hat{s})} =\kappa^2_{g,\text{off-shell}}(\hat{s}) \cdot \kappa^2_{V,\text{off-shell}}(\hat{s}) \quad ,
\end{equation}
\end{linenomath}
where $\kappa_{g,\text{off-shell}}(\hat{s})$ and $\kappa_{V,\text{off-shell}}(\hat{s})$ are the 
off-shell coupling scale factors associated with the $gg\to H^*$ production and the $H^*\to VV$ decay.
Due to the statistically limited sensitivity of the current analysis, the off-shell signal strength and 
coupling scale factors are assumed in the following to be independent of $\hat{s}$ in the high-mass region selected by the analysis.
The off-shell Higgs boson signal cannot be treated independently from the $gg \to VV$ background, 
as sizeable negative interference effects appear~\cite{Kauer:2012hd}. 
The interference term is proportional to $\sqrt{\mu_\text{off-shell}}=\kappa_{g,\text{off-shell}} \cdot \kappa_{V,\text{off-shell}}$.

%%%%%%%%%%%%%%%% 
%%%%%%%%%%%%%%%% 

In contrast, the cross-section for on-shell Higgs boson production allows a measurement of the signal strength:
\begin{linenomath}
\begin{equation}
  \mu_\text{on-shell} \equiv \frac{\sigma_\text{on-shell}^{gg \to H \to VV}}{\sigma_\text{on-shell, SM}^{gg \to H \to VV}}=\frac{\kappa^2_{g,\text{on-shell}} \cdot \kappa^2_{V,\text{on-shell}}}{\Widthratio}, 
\end{equation}
\end{linenomath}
which depends on the total width $\Gamma_H$.
Assuming identical on-shell and off-shell Higgs boson coupling scale factors,
the ratio of $\muoffshell$ to $\muonshell$ provides a measurement of the total width of the Higgs boson.
This assumption is particularly relevant to the running of the effective coupling $\kappa_g(\hat{s})$ for the loop-induced 
$gg \to H$ production process, as it is sensitive to new physics that enters at higher mass scales and 
could be probed in the high-mass $m_{VV}$ signal region of this analysis.
More details are given in Refs.~\cite{Englert:2014aca,Cacciapaglia:2014rla,Azatov:2014jga,Ghezzi:2014qpa,Buschmann:2014sia}.
With the current sensitivity of the analysis, only an upper limit on the total width $\Gamma_H$ can be determined, 
for which the weaker assumption 
\begin{linenomath}
\begin{equation}
\label{eqn:kappa_onshell<offshell}
\kappa^2_{g,\text{on-shell}} \cdot \kappa^2_{V,\text{on-shell}} \leq \kappa^2_{g,\text{off-shell}} \cdot \kappa^2_{V,\text{off-shell}} \, , 
\end{equation}
\end{linenomath}
that the on-shell couplings are no larger than the off-shell couplings, is sufficient.
It is also assumed that any new physics which modifies the off-shell signal strength $\mu_\text{off-shell}$ and 
the off-shell couplings $\kappa_{i,\text{off-shell}}$ does not modify the predictions 
for the backgrounds.
Further, neither are there sizeable kinematic modifications to the off-shell signal nor new, sizeable signals in the search region of this analysis unrelated to an enhanced off-shell signal strength~\cite{Englert:2014ffa,Logan:2014ppa}.

While higher-order quantum chromodynamics (QCD) and electroweak (EW) corrections are known for the off-shell signal process 
$gg \to H^* \to ZZ$~\cite{Passarino:2013bha}, which are also applicable to $gg \to H^* \to WW$, 
no higher-order QCD calculations are available for the $gg \to VV$ background process,
which is evaluated at leading order (LO).
Therefore the results are given as a function of the unknown K-factor for the $gg\to VV$ background.
QCD corrections for the off-shell signal processes have only been calculated inclusively in the jet multiplicity.
The experimental analyses are therefore performed inclusively in jet observables and,
the event selections are designed to minimise the dependence on the boost of the $VV$ system, 
which is sensitive to the jet multiplicity.

The dominant processes contributing to the high-mass signal region in the $ZZ \to 4\ell$, $ZZ \to 2\ell2\nu$ and $\ww$ final states are: 
the $gg \to H^* \to VV$ off-shell signal, the $gg \to VV$ continuum background, the interference between them, 
$VV$ production in association with two jets through VBF and $VH$-like production modes $pp\to VV+2j$ ($s$-, $t$- and $u$-channel) 
and the $q\bar{q} \to VV$ background. The LO Feynman diagrams for the $gg \to H^* \to VV$ signal, 
the continuum $gg \to VV$ background and the dominant irreducible $q \bar{q} \to VV$ background are depicted in Fig.~\ref{fig:feynman_diagrams}.
The $\ww$ channel also receives sizeable background contributions from $t\bar{t}$ and 
single-top production.
In the following a Higgs boson mass of $m_H=125.5$~\GeV, close to the ATLAS-measured Higgs boson mass value 
of $125.36$~\GeV~\cite{Aad:2014aba}, is assumed for the off-shell signal processes. This small difference has a negligible impact 
on the predicted off-shell production yields.

\begin{figure}[!ht]
\begin{center}
\subfigure[]{\includegraphics[width=0.55\figwidth]{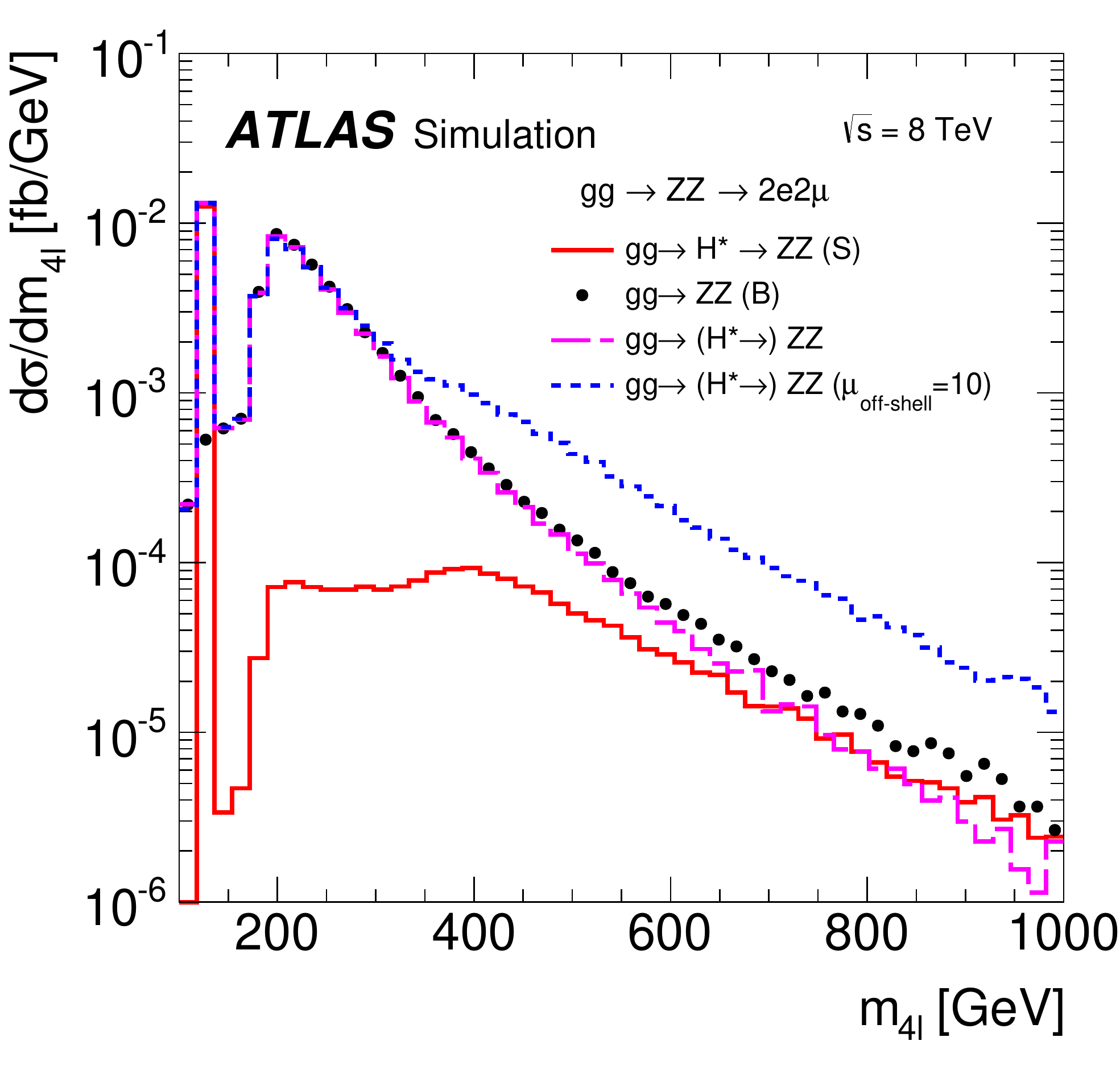}\label{fig:m4l_ggzz4l_lhelevel:m4l}}
\subfigure[]{\includegraphics[width=0.55\figwidth]{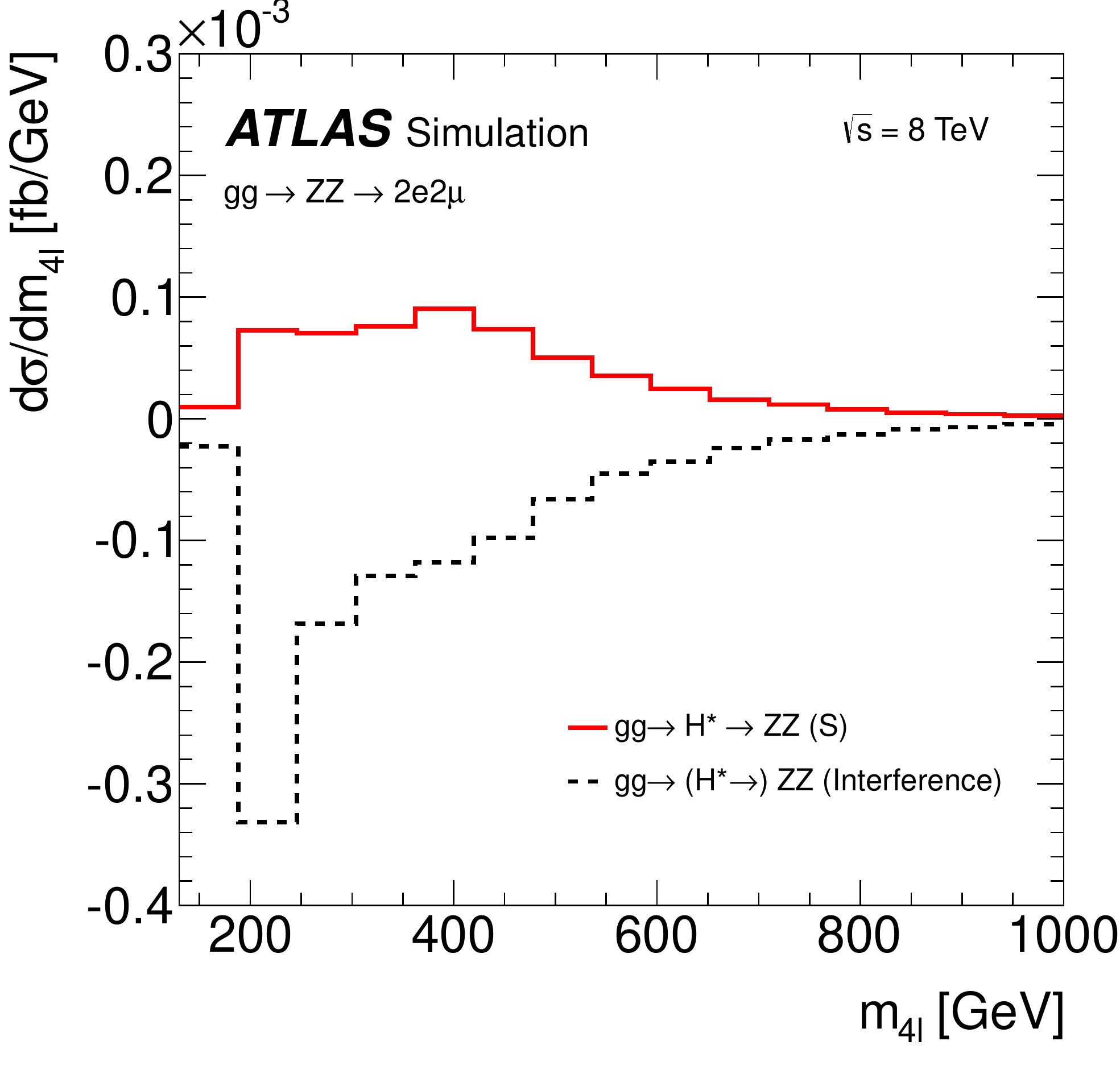}\label{fig:m4l_ggzz4l_lhelevel:interference}}
\caption{\subref{fig:m4l_ggzz4l_lhelevel:m4l} Differential cross-sections as a function of the four-lepton invariant mass
 $m_{4\ell}$ in the range of 100~\GeV~$<~m_{4\ell} <$~1000~\GeV\ for the $gg\to (H^*\to) ZZ\to 2e2\mu$ channel 
at the parton level, for the $gg\to H^*\to ZZ$ signal (solid line), $gg\to ZZ$ continuum background (dots), 
$gg\to (H^*\to) ZZ$ with SM Higgs boson coupling (long-dashed line, including signal plus background plus interference) and $gg\to (H^* \to) ZZ$ with $\mu_\text{off-shell}=10$ (dashed line). 
\subref{fig:m4l_ggzz4l_lhelevel:interference} Differential cross-section as a function of 
$m_{4\ell}$ in the range of 130~\GeV~$< m_{4\ell} <$~1000~\GeV\ for the SM $gg\to H^*\to ZZ\to 2e2\mu$ signal (solid line) and its interference 
with the $gg\to ZZ\to 2e2\mu$ continuum background (dashed line). 
}
\label{fig:m4l_ggzz4l_lhelevel}
\end{center}
\end{figure}

Figure~\ref{fig:m4l_ggzz4l_lhelevel} illustrates the size and kinematic properties of the gluon-induced signal and background processes
by showing the four-lepton invariant mass ($m_{4\ell}$) distribution for the $gg \to (H^*\to) ZZ\to 2e2\mu$ processes
after applying the event selections in the $ZZ\to 4\ell$ channel (see Sect.~\ref{sec:4l}) on generator-level quantities. 
The process $gg \to (H^*\to) ZZ\to 2e2\mu$ is shown for the SM $\mu_\text{off-shell}=1$ case and for an increased off-shell signal with  $\mu_\text{off-shell}=10$.
For low masses $m_{ZZ}<2m_Z$ the off-shell signal is negligible, 
while it becomes comparable to the 
continuum $gg \to ZZ$ background for masses above the $2m_t$ threshold. 
The interference between the $gg \to H^* \to ZZ$ signal and the $gg\to ZZ$ background is negative 
over the whole mass range.
A very similar relation between the $gg \to H^* \to VV$ signal and the $gg\to VV$ background is also seen for the 
$gg \to (H^*\to) ZZ\to 2\ell 2\nu$ and $gg \to (H^*\to) \ww$ processes.
 
%%%%%%%%%%%%%%%% 
%%%%%%%%%%%%%%%% 

The detector simulation for most generated Monte Carlo (MC) event samples is performed using Geant4~\cite{GEANT4,atlas-sim}.
Some background MC samples in the $\ww$ analysis for processes with large cross-sections are simulated with 
the fast detector simulation package Atlfast-II~\cite{atlas-sim}.

\subsection{Simulation of $gg \to (H^* \to) VV$}
\label{sec:gg2vvsim}
To generate the $gg \to H^* \to VV$ and $gg\to VV$ processes, including the interference, 
the LO MC generators gg2VV~\cite{Kauer:2012hd,Kauer:2013qba} and MCFM~\cite{Campbell:2013una,Campbell:2013wga} 
together with PYTHIA8~\cite{Sjostrand:2007gs}
and SHERPA+OpenLoops~\cite{Cascioli:2013gfa,Gleisberg:2008ta,Cascioli:2011va,Denner:2014gla} are used. 
The QCD renormalisation and factorisation scales are set to $m_{VV}/2$~\cite{Campbell:2013una}. 
The CT10 next-to-next-to-leading-order (NNLO) PDF set~\cite{Gao:2013xoa} is used, as the LO $gg\to VV$ process is part of the NNLO calculation for $pp \to VV$. 
The default parton showering and hadronisation option for the events processed with the full detector simulation 
is PYTHIA8 with the ``power shower'' parton shower option~\cite{Sjostrand:2007gs}.

For the $gg\to H^*\to VV$ signal, a NNLO/LO K-factor\footnote%
{The shorter $gg \to X$ notation is used also in the context of higher-order QCD calculations 
where $qg$ and $qq$ initial states contribute to the full $pp \to X$ process.} including the next-to-leading-order (NLO) electroweak corrections,
$\KHmVV=\sigma^\text{NNLO}_{gg\to H^*\to VV}/\sigma^\text{LO}_{gg\to H^*\to VV}$, is applied.
The K-factor and associated uncertainties are calculated in Ref.~\cite{Passarino:2013bha}
as a function of the Higgs boson virtuality $m_{VV}$ for $m_H\sim$125.5~\GeV, using the MSTW2008 PDF set~\cite{Martin:2009iq}.
Additional corrections are used to re-weight the predictions to the CT10 NNLO PDF set used in the simulation.

For the $gg\to VV$ background and the interference with the $gg\to H^*\to VV$ signal,
no higher-order QCD calculations are available.
However, these corrections are studied for the $WW$ final state in Ref.~\cite{Bonvini:2013jha} in 
the soft-collinear approximation, which is considered suitable for high-mass Higgs boson production.
In this approximation, the signal K-factor is found to provide a reliable estimate for the 
higher-order QCD corrections to the signal-background interference term.

The K-factor for the $gg\to VV$ background process, $\text{K}(gg\to VV)$, remains unknown.
Therefore, the results in this note are given as a function of the unknown K-factor ratio 
between the $gg \to VV$ background and the $gg\to H^*\to VV$ signal, defined as
\begin{linenomath}
\begin{equation}
\Kratio=\frac{\text{K}(gg \to VV)}{\text{K}(gg \to H^* \to VV)}=\frac{\KBmVV}{\KHggmVV} \, , 
\end{equation}
\end{linenomath}
where \KBmVV\ is the unknown mass-dependent K-factor for the $gg \to VV$ background, and
\KHggmVV\ is the gluon-initiated K-factor~\cite{Passarino:2013bha} for the signal\footnote{%
Numerically, 
\KHggmVV\ differs from \KHmVV\ by $\sim 2\%$ as the higher-order QCD contribution from $qg$ and $qq$ production is small.
However, \KHggmZZ\ has substantially larger uncertainties than \KHmZZ. Therefore \KHmZZ\ is substituted here, 
ignoring the 2\% shift in central value, but taking the difference in the systematic uncertainty into account.}
as motivated by the soft-collinear approximation in Ref.~\cite{Bonvini:2013jha}.
Because the K-factor \KHggmVV\ changes by less than 10\% as a function of $m_{VV}$ in the relevant region of phase space, 
no mass dependence on \Kratio\ is assumed.
The range 0.5--2 is chosen for the variation of the K-factor ratio \Kratio\ in order to include 
the full correction from the signal K-factor $\KHggmVV\sim 2$ in the variation range. 
With respect to the LO $gg \to VV$ process,
this corresponds to an absolute variation in the approximate range 1--4.
Using the K-factors discussed above, the cross-section
for the $gg \to {(H^* \to)}VV$ process with any off-shell Higgs
boson signal strength $\mu_\text{off-shell}$ can be parameterised as:
%%%%%%%%%%%%%%%%%%%%
\begin{eqnarray}
\label{eqn:sigma_ggHVVscaling_on_S_B_I}
\sigma_{gg\rightarrow (H^* \rightarrow) VV}(\mu_\text{off-shell},m_{VV})& = & 
\KHmVV \cdot \mu_\text{off-shell} \cdot \sigma_{gg\rightarrow H^*\rightarrow VV}^\text{SM}(m_{VV}) \\\nonumber
& + & \sqrt{ \KHggmVV \cdot \KBmVV \cdot \mu_\text{off-shell} } \cdot \sigma_{gg\rightarrow VV,\,\text{Interference}}^{\text{SM}}(m_{VV}) \\\nonumber
& + & \KBmVV \cdot \sigma_{gg\rightarrow VV,\, \text{cont}}(m_{VV}) \,.
\end{eqnarray}
More details are given in Appendix~\ref{Sec:App:ggHVVscaling}.

In addition, higher-order QCD corrections to the transverse momentum%
\footnote{ATLAS
  uses a right-handed coordinate system with its origin at the nominal
  interaction point (IP) in the centre of the detector, and the
  $z$-axis along the beam line.  The $x$-axis points from the IP to
  the centre of the LHC ring, and the $y$-axis points upwards.
  Cylindrical coordinates $(r,\phi)$ are used in the transverse plane,
  $\phi$ being the azimuthal angle around the beam line.  Observables
  labelled ``transverse'' are projected into the $x\text{--}y$ plane.
  The pseudorapidity is defined in terms of the polar angle $\theta$ as
  $\eta=-\ln\tan(\theta/2)$.}
\pt\ and the rapidity $y$ of the $VV$ system are studied using SHERPA+OpenLoops,
which includes matrix-element calculations for the first hard jet emission.
A difference of order 20\% in the ratio of the \pt\ of the $VV$ system in the relevant kinematic region 
is observed when comparing the LO generators with parton shower to SHERPA+OpenLoops, while the difference in the 
rapidity $y$ of the $VV$ system is small.
This difference in the \pt\ of the $VV$ system can modify the kinematic observables used in the analyses,
leading to variations in both the kinematic shapes and acceptance which are not
covered by the $m_{VV}$ dependent systematic uncertainties derived in Ref.~\cite{Passarino:2013bha}.
To account for these effects, the LO generators are re-weighted to
SHERPA+OpenLoops in the \pt\ of the $VV$ system.
Due to the different jet emission mechanisms in the signal
and the background processes, different re-weighting functions are derived for the
$gg\to H^*\to VV$ signal, the $gg\to VV$ background, and the total $gg\to (H^*)\to VV$, respectively.
The impact of the re-weighting on the acceptance is below 1\% for the signal
and at the level of 4--6\% for the background.
In the $ZZ\to 4\ell$ channel, the re-weighting procedure is only used to
account for the acceptance effects, as the matrix-element-based discriminant is insensitive to
the \pt\ of the $ZZ$ system.
For the $ZZ\to 2\ell2\nu$ channel, the re-weighting is used in both
the transverse mass shape and acceptance as the $m_{\mathrm{T}}$ depends on the \pt\ of the $ZZ$ system.
For the $\ww$ channel, the re-weighting affects only the acceptance.

\subsection{Simulation of electroweak $VV$ production through VBF and $VH$-like processes}

The electroweak\footnote{Electroweak means in this context that QCD diagrams that enter through the QCD NNLO corrections 
to $pp\to VV$ are not included.} 
$pp\to VV + 2j$ processes contain both VBF-like events and $VH$-like events,
which are simulated using MadGraph5~\cite{Alwall:2014hca} and cross-checked
using PHANTOM~\cite{Ballestrero:2007xq}.
The QCD renormalisation and factorisation scales are set to $m_W$ following the recommendation 
in Ref.~\cite{LHCHiggsCrossSectionWorkingGroup:2012vm} and the CTEQ6L1 PDF set~\cite{Pumplin:2002vw} is used. 
PYTHIA6~\cite{Sjostrand:2006za} is used for parton showering and hadronisation. 

The high-mass range selected by this analysis includes Higgs boson signal events arising from:
\begin{itemize}
 \item the off-shell VBF $H \to VV$ process, which scales with $\kappa^4_{V,\text{off-shell}}$ and is independent of $\Gamma_H$,
 \item VBF-like $VV$ processes with a $t$-channel Higgs boson exchange, which scale with $\kappa^4_{V,\text{off-shell}}$ and are independent of $\Gamma_H$,
 \item $WH$ and $ZH$ processes with an on-shell Higgs boson, with decays $Z \to 2\ell$ or $W \to \ell\nu$ and $H \to 2\ell 2j$ or $H \to \ell\nu 2j$, which scale with $\kappa^4_{V,\text{on-shell}}/\Gamma_H$, 
\end{itemize}
where we assume the same coupling strength $\kappa_{V,\text{off-shell}}$ in the two VBF-like contributions, 
although the energy scale of the Higgs boson propagator is different between the two cases.
Due to the different $\Gamma_H$ dependence, the on-shell and off-shell Higgs boson production processes are 
separated in the analysis by requiring that the generated Higgs boson mass satisfies $|m_H^\text{gen.}-125.5\GeV|<1\GeV$.
This requirement is fully efficient in selecting the on-shell $VH$ process.
The NNLO QCD corrected cross-section in Ref.~\cite{Heinemeyer:2013tqa} is used for the on-shell $VH$ production process.
The cross-section $\sigma_{pp\rightarrow VV + 2j}(\mu_\text{off-shell})$ for the electroweak $pp\to VV + 2j$ process 
for any off-shell Higgs boson signal strength $\mu_\text{off-shell}$ is parameterised in 
the same way as for the $gg \to (H^* \to) VV$ process.
Details are given in Appendix~\ref{Sec:App:VBFHVVscaling}.

\subsection{Simulation of $q\bar{q} \to ZZ$, $WW$ and $WZ$ backgrounds} 
\label{sec:qqVV}

The $q\bar{q} \to ZZ$, $q\bar{q} \to WW$, and $q\bar{q} \to WZ$ backgrounds are simulated
at NLO in QCD using POWHEG-BOX~\cite{Melia:2011tj} with dynamic QCD renormalisation
and factorisation scales of $m_{VV^{(')}}$ and the CT10 NLO PDF set.
In addition, SHERPA is used as a cross-check for the $q{\bar q}\to WZ$ background.
Parton showering and hadronisation are done with PYTHIA8 for $q\bar{q} \to VZ$ and PYTHIA6 for $q\bar{q} \to WW$.
The interference between the $q{\bar q} \to WW$ and $q{\bar q} \to ZZ$ 
processes for the $2\ell2\nu$ final state is 
negligible~\cite{Melia:2011tj} and thus not considered.

The cross-sections for the $q\bar{q} \to ZZ$ and $q\bar{q} \to WW$ processes are calculated in Ref.~\cite{Cascioli:2014yka} 
and Ref.~\cite{Gehrmann:2014fva}, respectively, for two on-shell $Z$ or $W$ bosons 
in the final state at NNLO QCD accuracy. 
As these calculations include the $gg \to VV$ processes as part of the NNLO calculation,
a different K-factor is provided by the authors of the Refs.~\cite{Cascioli:2014yka,Gehrmann:2014fva}
excluding the $gg\to VV$ component and using a QCD renormalisation and factorisation scale  $\mu_{QCD }$ of $m_{VV}/2$ 
in order to consistently match the simulation of the $gg \to (H^* \to) VV$ process:
\begin{eqnarray}
\text{K}_{q \bar{q}}(m_{VV}) &=& \frac{\sigma_{q\bar{q} \to VV}^\text{NNLO}(m_{VV},\mu_{QCD}=m_{VV}/2) - \sigma_{gg \to VV}^\text{LO}(m_{VV},\mu_{QCD}=m_{VV}/2)}{\sigma_{q\bar{q} \to VV}^\text{NLO}(m_{VV},\mu_{QCD}=m_{VV})}.
\end{eqnarray}
Electroweak higher-order corrections are not included in POWHEG-BOX.
These corrections are calculated in Refs.~\cite{Bierweiler:2013dja,Baglio:2013toa}
for on-shell outgoing vector bosons and found to be about $-10\%$ in the high-mass
$VV$ region of this analysis.
To account for these corrections, the POWHEG-BOX events are re-weighted
using a procedure comparable to the one described in Ref.~\cite{Gieseke:2014gka}, 
based on the kinematics of the diboson system and the initial state quarks.

\subsection{Simulation of top-quark backgrounds} 
In the $\ww$ channel, 
the $t\bar{t}$ and single-top ($s$-channel and $Wt$) backgrounds are simulated with POWHEG-BOX~\cite{Alioli:2011as,Alioli:2009je} with parton 
showering and hadronisation done 
with PYTHIA6, using the CT10 NLO PDF set.
The $t$-channel single-top background is simulated using AcerMC\cite{Kersevan:2004yg}+PYTHIA6 and uses the CTEQ6LI PDF set.
The relative rates of $t\bar{t}$ and single-top production are evaluated with Top++2.0 \cite{Czakon20142930} and the calculations in Refs.~\cite{PhysRevD.81.054028,PhysRevD.83.091503,PhysRevD.82.054018} respectively.

%-------------------------------------------------------------------------------
\section{Analysis of the $ZZ \to 4\ell$ final state}
\label{sec:4l}
\label{sec:zz4l_sel_bkg}

The analysis for the $ZZ\to 4\ell$ final state closely follows the Higgs boson measurements in
the same final state described in Ref.~\cite{Aad:2014eva}, with the same physics object definitions,
trigger and event selections, and background estimation methods.
A matrix-element-based discriminant (ME-based discriminant)
is constructed to enhance the separation between the $gg\to H^*\to ZZ$ signal
and the $gg\to ZZ$ and $q{\bar q}\to ZZ$ backgrounds,
and is subsequently used in a binned maximum-likelihood fit for the final result.

\subsection{Event selection}
\label{sec:zz4lsel}

To minimise the dependence of the $gg\to ZZ$ kinematics on higher-order QCD effects, 
the analysis is performed inclusively, ignoring the number of jets in the events. 

The analysis is split into four lepton channels ($2\mu2e$, $2e2\mu$, $4e$, $4\mu$) as in Ref.~\cite{Aad:2014eva}.
Each electron (muon) must satisfy $E_{\mathrm{T}} > 7$~\GeV\ ($\pt > 6$~\GeV) and be measured in the
pseudorapidity range $|\eta| < 2.47$ ($|\eta| < 2.7$).
The highest-$\pt$ lepton in the quadruplet must satisfy $\pt > 20$~\GeV,
and the second (third) lepton in $\pt$ order must satisfy $\pt > 15$~\GeV\ ($\pt > 10$~\GeV).
Lepton pairs are formed from same-flavour opposite-charge leptons.
For each channel, the lepton pair with the mass closest to the $Z$ boson mass is
referred to as the leading dilepton pair and its invariant mass, $m_{12}$, is required to be between 50~\GeV\ and 106~\GeV.
The second (subleading) pair is chosen from the remaining leptons (more than four leptons are allowed per event) as the pair closest
in mass to the $Z$ boson and in the range of 50~\GeV~$ < m_{34} < 115$~\GeV.
The off-peak region is defined to include the range from 220~\GeV~$< m_{4\ell} <$~1000~\GeV.

Figure~\ref{fig:m4l_data_and_exp:m4l} shows the observed and expected
distributions of $m_{4\ell}$ combining all lepton channels in the full off-peak region. 
The data are in agreement with the SM predictions, with a small deficit of the order of one standard deviation ($1\sigma$).
Table~\ref{tab:yield_cut_data} shows the expected and observed number of events in
the signal-enriched region, 400~\GeV~$ <m_{4\ell}<$~1000~\GeV, combining all lepton channels. 
This mass region was chosen since it is optimal for a $m_{4\ell}$ cut-based analysis.

\subsection{Matrix-element-based kinematic discriminant}
\label{sec:mekd4l}

\begin{figure}[!htbp]
\begin{center}
\subfigure[]{\includegraphics[width=0.55\figwidth]{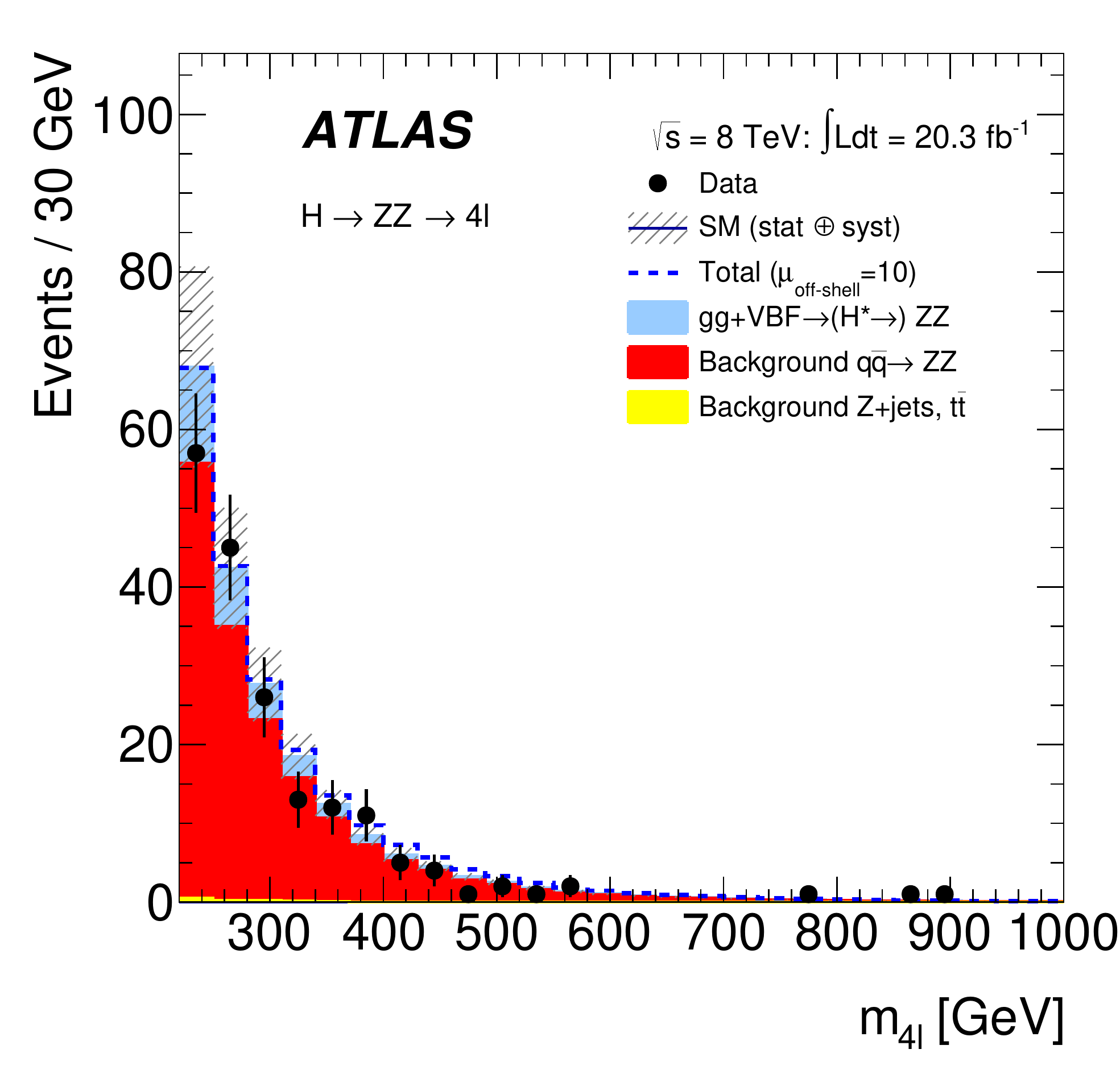}{\label{fig:m4l_data_and_exp:m4l}}}
\subfigure[]{\includegraphics[width=0.55\figwidth]{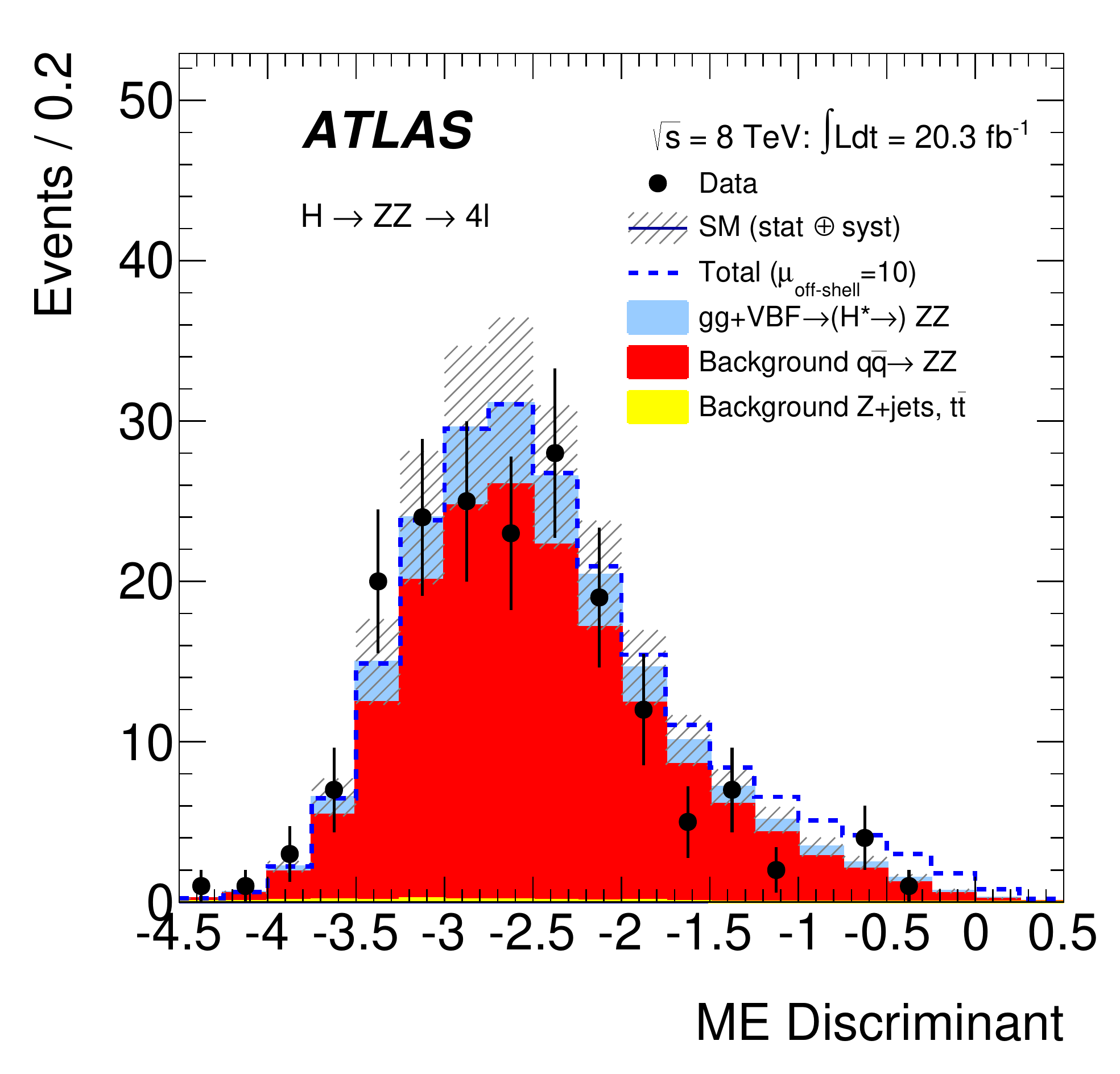}\label{fig:m4l_data_and_exp:ME}}
\caption{Observed distributions for \subref{fig:m4l_data_and_exp:m4l} the four-lepton invariant mass $m_{4\ell}$ in the range
of 220~\GeV~$< m_{4\ell} <$~1000~\GeV~ and \subref{fig:m4l_data_and_exp:ME} the ME-based discriminant
combining all lepton final states for the ME-based analysis signal region, compared to the expected contributions from the SM including the Higgs boson (stack). 
The dashed line corresponds to the total expected event yield, including all backgrounds and the Higgs boson with $\mu_\text{off-shell}=10$.
A relative $gg\to ZZ$ background K-factor of \Kratio=1.0 is assumed. 
The $Z$+jets and top-quark backgrounds are barely visible in the plot since they are very small (<1\% of the total background).
}
\label{fig:m4l_data_and_exp}
\end{center}
\end{figure}

The matrix-element kinematic discriminant fully exploits the event kinematics in the centre-of-mass frame of the $4\ell$ system, based on eight observables: 
\{$m_{4\ell}, m_{12}, m_{34}, \cos\theta_1, \cos\theta_2, \phi, \cos\theta^*, \phi_1$\}, 
defined in Refs.~\cite{Aad:2013xqa,Aad:2014eva}.
These observables are used to create the four-momenta of the leptons and incoming partons, which are then used to calculate matrix elements for different processes, provided by the MCFM program~\cite{Campbell:2013una}. The following matrix elements are calculated for each event in the mass range 220~\GeV~$< m_{4\ell} <$~1000~\GeV: 
\begin{itemize}
\item $P_{q\bar{q}}$: matrix element squared for the $q\bar{q}\to ZZ\to 4\ell$ process, 
\item $P_{gg}$: matrix element squared for the $gg\to (H^*\to) ZZ\to 4\ell$ process including the Higgs boson ($m_H=125.5$~\GeV) 
with SM couplings, continuum background and their interference,  
\item $P_H$: matrix element squared for the $gg\to H^*\to ZZ\to 4\ell$ process ($m_H=125.5$~\GeV).
\end{itemize}
The kinematic discriminant is defined as in Ref.~\cite{Campbell:2013una}:
\begin{linenomath}
\begin{equation}
\label{eq:mekd}
\text{ME} = \log_{10}\left({\frac{P_H}{P_{gg}+c\cdot P_{q\bar{q}}}} \right), 
\end{equation}
\end{linenomath}
where $c=0.1$ is an empirical constant, to approximately balance the overall cross-sections of the $q\bar{q}\to ZZ$ and $gg\to (H^*\to) ZZ$ processes. 
The value of $c$ has a very small effect on the analysis sensitivity.
Figure~\ref{fig:m4l_data_and_exp:ME} shows the observed and expected
distributions of the ME-based discriminant combining all lepton final states.
Events with the ME-based discriminant value between $-$4.5 and 0.5 are selected with
a signal efficiency of $>$~99\%.

%%%%%%%%%%%%%%%% 
%%%%%%%%%%%%%%%% 

In addition, an alternative multivariate discriminant based on a boosted decision tree (BDT) algorithm was studied
to further separate the $gg\to H^*\to ZZ$ signal and the main $q{\bar q}\to ZZ$ background, 
by exploiting additional kinematic information ($\pt$ and $\eta$) of the $ZZ$ system.
The analysis sensitivity improves very little ($\sim$2\%) compared to the ME-based discriminant alone.
Due to the dependence on the $\pt$ of the $ZZ$ system, the BDT-based discriminant
introduces additional systematic uncertainties from the higher-order QCD corrections.
For these reasons, the BDT-based discriminant is not used for the final result.

%%%%%%%%%%%%%%%%%%%%%%%%%%%%%
\begin{table}[!ht]
\begin{center}
\begin{tabular}{|l|c|c|c|}
\hline
Process & $ZZ\to 4\ell$ & $\zzn$ & $\ww$ \\ 
\hline
$gg\to H^*\to VV$ (S) &  1.1 $\pm$  0.3  & 3.2 $\pm$ 1.0 & 1.5 $\pm$ 0.4 \\
$gg\to VV$ (B)        &  2.8 $\pm$  0.8  & 5.3 $\pm$ 1.6 &3.6 $\pm$ 1.1   \\
{\bf $gg\to (H^*\to) VV$}   &  {\bf 2.4 $\pm$  0.7}  & \bf{ 3.9 $\pm$ 1.2} & {\bf 2.4 $\pm$ 1.2} \\
$gg\to (H^*\to) VV$ ($\mu_\text{off-shell}=10$) & 9.2 $\pm$  2.5  & 24.0 $\pm$ 7.3 & 10 $\pm$ 4\\
\hline
VBF $H^*\to VV$ (S)	& 0.12 $\pm$  0.01  & 0.48 $\pm$ 0.04 & 0.42 $\pm$ 0.05\\
VBF $VV$ (B)            & 0.71 $\pm$  0.04  & 1.2 $\pm$ 0.2 &     1.6 $\pm$ 0.2   \\
{\bf VBF $(H^*\to)VV$}    & {\bf 0.59 $\pm$  0.03}  & {\bf 0.7 $\pm$ 0.1} &     {\bf 1.1 $\pm$ 0.1}  \\
VBF $(H^*\to)VV$ ($\mu_\text{off-shell}=10$) & 1.17 $\pm$  0.06  & 2.9 $\pm$ 0.2 & 2.8 $\pm$ 0.3  \\
\hline
$q{\bar q}\to ZZ$ & 21.3 $\pm$  2.1  & 31.5 $\pm$ 3.5 & \multirow{2}{*}{{$\left\}\rule{0cm}{5mm}\right.$ } 2.0 $\pm$ 0.2} \\
$q\bar{q}\to WZ$ & -  & 10.6 $\pm$ 1.4 &  \\
$q\bar{q}\rightarrow WW$ & -  & \multirow{3}{*}{{$\left\}\rule{0cm}{7mm}\right.$} 0.4 $\pm$ 0.2} & 40 $\pm$ 5 \\
$t\bar{t}$, $Wt$, and $t\bar{b}/tq\bar{b}$ & -  &  & 35 $\pm$ 4 \\
$Z\to\tau\tau$ & -  & & 1.4 $\pm$ 0.2 \\
$Z\to ee, \mu\mu$ & -  &  3.5 $\pm$ 3.0 &  - \\ 
Other backgrounds &  -   & 0.8 $\pm$ 0.2 & 8.7 $\pm$ 1.3 \\
\hline
Total Expected (SM) & 24.4 $\pm$  2.2  & 51 $\pm$ 6 & 90 $\pm$ 4\\
Observed  &   18 & 48 & 82\\ \hline
\end{tabular}
\caption{Expected and observed numbers of events in the signal region
for all final states in the cut-based approaches. 
For the $ZZ\to 4\ell$ analysis a mass range of $400<m_{4\ell}<1000$~\GeV\ is used.
The other backgrounds in the $ZZ\to 4\ell$ final state include contributions from $Z$+jets and top-quark processes.  
For the $ZZ\to 2\ell2\nu$ analysis the range 380~\GeV~$< m_{\mathrm{T}}^{ZZ} <$~1000~\GeV\ is considered.
For the $\ww$ analysis, the region R$_8>450$~\GeV\ is used and background event yields are quoted after the likelihood fit was performed.
The expected events for the $gg\to (H^*\to) VV$ and VBF $(H^*\to)VV$ processes ($ZZ$ or $WW$), including the Higgs boson signal, background and interference, are reported 
for both the SM predictions (in bold) and $\mu_\text{off-shell} = 10$.
A relative $gg\to VV$ background K-factor of \Kratio=1 is assumed.
The uncertainties in the number of expected events include the statistical 
uncertainties from MC samples and systematic uncertainties.
The entries with a $-$ are for processes with event yields $<0.1$.
}
\label{tab:yield_cut_data}
\end{center}
\end{table}
%%%%%%%%%%%%%%%%%%%%%%%%%%%%%

%-------------------------------------------------------------------------------
\section{Analysis of the $ZZ \to 2\ell\,2\nu$ final state}
\label{sec:2l2n}
The analysis of the $ZZ\to 2\ell 2\nu$ channel
follows strategies similar to those used in the invisible Higgs boson
search in the $ZH$ channel~\cite{Aad:2014iia}.  
The definitions of the reconstructed physics objects (electrons, muons, jets, 
and missing transverse momentum) are identical, but some of the kinematic cuts 
were optimised for the current analysis.

\subsection{Event selection} 
As the neutrinos in the final state do not allow for a kinematic reconstruction of $m_{ZZ}$, 
the transverse mass ($m_{\mathrm{T}}^{ZZ}$) reconstructed from the transverse momentum of the 
dilepton system ($p_{\mathrm{T}}^{\ell\ell}$)
and the magnitude of the missing transverse momentum (\met):
\begin{linenomath}
\begin{equation}\label{eq:mt}
m_{\mathrm{T}}^{ZZ} \equiv \sqrt{ 
        \left( \sqrt{m_{Z}^2+\left|\bm{p}_{\mathrm{T}}^{\ell\ell} \right|^2} +
               \sqrt{m_{Z}^2+\left|\bm{E}_{\mathrm{T}}^{\mathrm{miss}}\right|^2} \right)^2 
- 
\left|\bm{p}_{\mathrm{T}}^{\ell\ell}+ \bm{E}_{\mathrm{T}}^{\mathrm{miss}}\right|^2\ },
\end{equation}
\end{linenomath}
is chosen as the discriminating variable to enhance sensitivity to the $gg \to H^* \to ZZ$ signal. 

The selection criteria are optimised to maximise the signal significance with respect to the 
main backgrounds, which are $ZZ$, $WZ$, $WW$, top-quark, and $W/Z$+jets events, as described 
in Sect.~\ref{sec:bg2l2v}. 
The impact of the background uncertainty is considered in the significance calculation. 
 
First, events with two oppositely charged electron or muon candidates in the 
$Z$ mass window 76~\GeV~$< m_{\ell\ell} <$~106~\GeV~are selected. 
Events with a third lepton ($e$ or $\mu$) identified using 
looser identification criteria for the electrons and a lower \pt~threshold of 
7~\GeV~are rejected. 
A series of selection requirements are necessary to suppress the Drell--Yan background, including: \met $>$ 180~\GeV; 380~\GeV~$< m_{\mathrm{T}}^{ZZ} <$ 1000~\GeV; 
the azimuthal angle between the transverse momentum of the dilepton system and the missing transverse momentum 
\begin{math}
\Delta \phi(\pt^{\ell\ell},\met) > 2.5
\end{math}; and 
\begin{math}
\Bigl| \bigl|\bm{E}_{\mathrm{T}}^{\mathrm{miss}}+\sum_{\mathrm{jet}}\bm{p}_{\mathrm{T}}^{\mathrm{jet}} \bigl| - \pt^{\ell\ell} \Bigl|/\pt^{\ell\ell} < 0.3
\end{math}.
Events with a $b$-jet with \pt $>$ 20~\GeV\ and $|\eta|<$ 2.5,
identified by the MV1 algorithm~\cite{ATLAS-CONF-2014-004,ATLAS-CONF-2014-046}
with 70\% tagging efficiency, are rejected to suppress the top-quark background. 
Finally, the selection on the azimuthal angle between the two leptons $\Delta \phi_{\ell \ell} < 1.4$ is applied to select events with boosted $Z$
bosons to further discriminate the signal from the background.

\subsection{Background estimation}
\label{sec:bg2l2v}

The dominant background is $q\bar{q} \rightarrow ZZ$ production, 
followed by $q\bar{q} \rightarrow WZ$ production. 
Background contributions from events with a genuine isolated 
lepton pair, not originating from a
$Z\rightarrow ee$ or $Z\rightarrow \mu\mu$ decay, 
arise from the 
$WW$, $t\bar{t}$, $Wt$, and $Z\to\tau\tau$ processes. 
The remaining backgrounds are from $Z\rightarrow ee$ or 
$Z\rightarrow \mu\mu$ decays with poorly reconstructed \met, and 
from events with at least one misidentified electron or muon 
coming from $W$+jets, semileptonic top decays ($t\bar{t}$ and 
single top), and multi-jet events. 

The $q\bar{q} \rightarrow ZZ$ background is estimated in the same way
as for the $ZZ \to 4\ell$ analysis using the POWHEG-BOX simulation as 
described in Sect.~\ref{sec:qqVV}.
The $WZ$ background is also estimated with the simulation 
(described in Sect.~\ref{sec:qqVV}) and validated with data in a 
three-lepton control region. 
The observed number of events in the control region for \met$>$ 180~\GeV (300~\GeV) is 
30 (3), whereas the predicted event
yield is 22.9 $\pm$ 0.8 (3.4 $\pm$ 0.3). No significant difference
is observed between the data and simulation.   

The $WW$, $t\bar{t}$, $Wt$, and $Z\to\tau\tau$ backgrounds are 
inclusively estimated with data assuming lepton flavour symmetry in 
an $e\mu$ control region using a relaxed selection. 
The following equations show how these backgrounds in the signal 
region can be estimated with $e\mu$ events: 
\begin{eqnarray} 
\label{eq:ee_emubkg}
\begin{split}
N^{\rm bkg}_{ee} &=& \frac{1}{2} \times N^{\rm data,sub}_{e\mu}\times{\alpha}, \\
N^{\rm bkg}_{\mu\mu} &=& \frac{1}{2} \times N^{\rm data,sub}_{e\mu}\times\frac{1}{\alpha},
\end{split}
\end{eqnarray}
where $N^{\rm bkg}_{ee}$ and $N^{\rm bkg}_{\mu\mu}$ are the number of dielectron and dimuon events 
in the signal region.
$N^{\rm data,sub}_{e\mu}$ is the number of events in the $e\mu$ 
control region with 
$WZ$, $ZZ$, and other small backgrounds ($W$+jets, $t\bar{t}V$, and triboson)
subtracted using simulation.  
The different $e$ and $\mu$ efficiencies are taken into account using the $\alpha$ variable, 
which is an efficiency correction factor determined from 
the ratio of dielectron to dimuon event yields after the inclusive $Z$ mass requirement 
(76~\GeV~$< m_{\ell\ell} <$~106~\GeV). 
The measured value of $\alpha$ is 0.942 with a systematic uncertainty 
of 0.004 and a negligible statistical uncertainty. 
This scale factor is applied to the MC predictions.
The other source of systematic uncertainty comes 
from the subtraction of 
$WZ$, $ZZ$, and other small backgrounds 
in the $e\mu$ control region using the simulation. 
As no data event remains after applying the full selection, a scale 
factor of 1.4 $\pm$ 0.3 is derived by comparing the event yields
from the data-driven and MC predictions with a relaxed selection 
applying the \met\ and $m_{\mathrm{T}}^{ZZ}$~requirements but no further cuts.
Experimental systematic uncertainties are considered for the 
MC predictions.

Imperfect modelling of detector non-uniformities and \met~response could lead to 
an incorrect estimate of the $Z$ boson background in the signal region.
The $Z$ boson background is estimated with data using the two-dimensional sideband
regions constructed by reversing one or both of the 
$\Delta \phi(\pt^{\ell\ell}$,\met) and $\Delta \phi_{\ell \ell}$ 
selections~\cite{Aad:2014iia}.
The main uncertainty on the mis-measured $Z$ boson background arises
from the differences in shape of the \met\ and $m_{\mathrm{T}}^{ZZ}$
distributions in the signal and sideband regions and the small 
correlation between these two variables. 
Other systematic uncertainties originate from
the subtraction of the non-$Z$ boson backgrounds in the sideband regions.

The $W$+jets and multi-jet backgrounds are estimated from data using 
the fake-factor method~\cite{Aad:2014iia}. The 
predicted background with a looser \met~selection applied at 
100~\GeV, and without the $m_{\mathrm{T}}^{ZZ}$ selection, is 0.04 $\pm$ 0.01 events. 
No event remains after applying the full event selection
for both the data-driven method and MC samples, and 
hence this background is estimated to be negligible. 

\begin{figure}[!htbp]
\begin{center}
\includegraphics[width=0.55\figwidth]{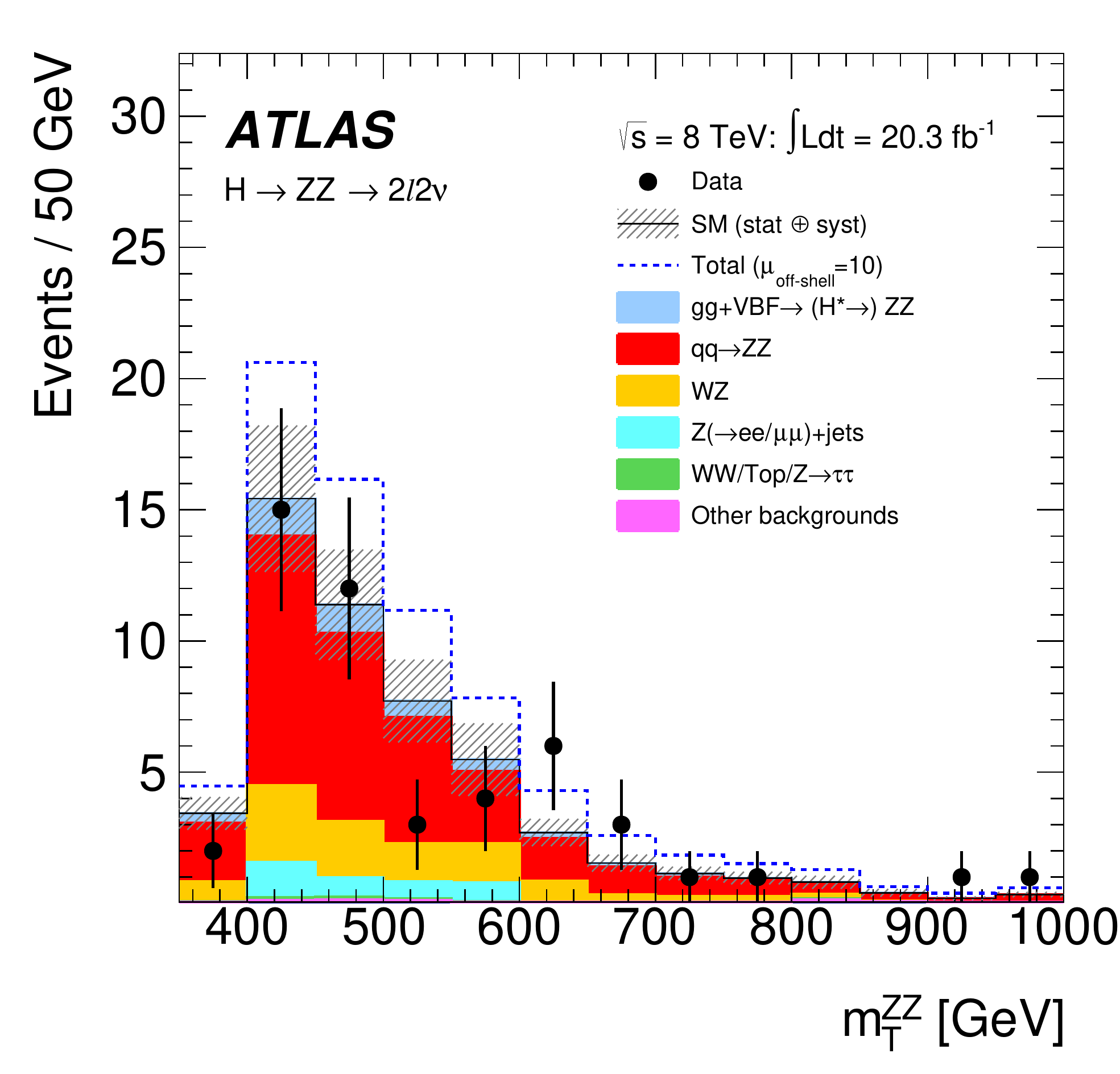}
\caption{Observed distribution of the $ZZ$ transverse mass $m_{\mathrm{T}}^{ZZ}$ 
in the range 380~\GeV~$< m_{\mathrm{T}}^{ZZ} <$~1000~\GeV~combining  
the $2e2\nu$ and $2\mu2\nu$ channels, compared to the expected 
contributions from the SM including the Higgs boson (stack).
The first bin only contains events in the range 380~\GeV~$< m_{\mathrm{T}}^{ZZ} <$~400~\GeV.
The hatched area shows the combined statistical and 
systematic uncertainties. 
The dashed line corresponds to the total expected event yield, including 
all backgrounds and the Higgs boson with $\mu_\text{off-shell}=10$.
A relative $gg\to ZZ$ background K-factor of \Kratio=1 is assumed.}
\label{fig:zz2l2v_sr_mt}
\end{center}
\end{figure}

The predicted signals and backgrounds with statistical and systematic
uncertainties are summarised in Table~\ref{tab:yield_cut_data}. 
The observed event yields agree with the total predicted ones from the 
SM within the uncertainties. 
Figure~\ref{fig:zz2l2v_sr_mt} shows the distributions of $m_{\mathrm{T}}^{ZZ}$ for the $ee$ and $\mu\mu$ channels 
in the signal region, compared to the predicted contributions from the SM as well as to a 
Higgs boson with $\mu_\text{off-shell} = 10$.

%-------------------------------------------------------------------------------
\section{Analysis of the $WW \to e\nu\,\mu\nu$ final state}
\label{sec:lnln}
The analysis of the $\ww$ channel closely follows the Higgs boson measurements
in the oppositely charged electron--muon pair final state in Ref.~\cite{ATLAS:2014aga}.  
This selection ensures orthogonality with the $\zzn$ final state.  
The same object identification and selection as in  Ref.~\cite{ATLAS:2014aga} is used in this analysis.  
Additionally, an event selection identical to that used for the gluon fusion initial states in $H\rightarrow \ww$ is used, up to and including a requirement on missing transverse momentum: leading lepton $\pt>22$~\GeV, subleading lepton $\pt>10$~\GeV, $m_{\ell\ell}>10$~\GeV, and $\pt^{\text{miss,track}}>20$~\GeV, the magnitude of the missing transverse momentum, with a track-based soft term.  
The signal region (SR) and background estimations were revised for the high-mass region used in this analysis.
Contrary to the base analysis \cite{ATLAS:2014aga}, events are not binned by the number of jets.  
Top-quark events and SM~$WW$ production remain the largest expected backgrounds.  

\subsection{Event selection}

\label{sec:hww_event_selection}

As with the $\zzn$ channel, the neutrinos in the final state do not allow for a kinematic reconstruction of $m_{VV}$.  Thus a transverse mass $(m_{\mathrm{T}}^{WW})$ is calculated from the dilepton system transverse energy $(E_{\text{T}}^{\ell\ell})$, the vector sum of lepton transverse momenta $(\boldsymbol{p}_{\text{T}}^{\ell\ell})$, and the vector sum of neutrino transverse momenta $(\boldsymbol{p}_{\text{T}}^{\nu\nu})$, measured with $\pt^{\text{miss,track}}$:
\begin{linenomath}
\begin{equation}
  m_{\mathrm{T}}^{WW} = \sqrt{\left(E_{\text{T}}^{\ell\ell} + \pt^{\nu\nu}\right)^2 - \left|\boldsymbol{p}_{\text{T}}^{\ell\ell} + \boldsymbol{p}_{\text{T}}^{\nu\nu}\right|^2}\mathrm{,~where }~E_{\text{T}}^{\ell\ell}=\sqrt{\big(\pt^{\ell\ell}\big)^2+\big(m_{\ell\ell}\big)^2}.
\label{eqn:mTHww}
\end{equation}
\end{linenomath}
The transverse mass is modified compared to the definition in Eq.~(\ref{eq:mt}) as the neutrinos do not come from the same parent particle, and there is no $m_{Z}$ constraint.

\begin{figure}[htbp]
\centering
\subfigure[]{\includegraphics[width=0.55\figwidth]{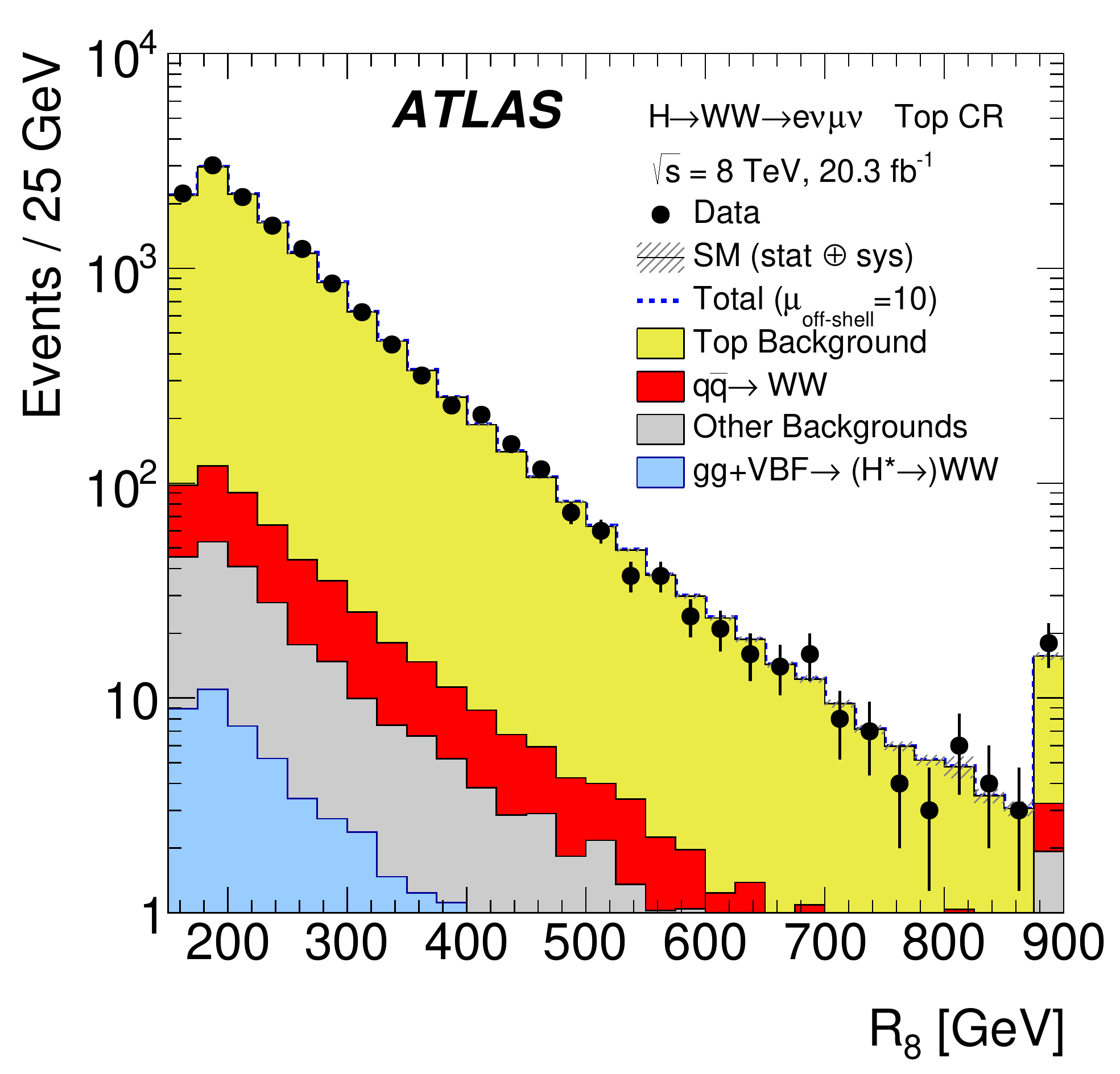}\label{fig:hww_topcr}}
\subfigure[]{\includegraphics[width=0.55\figwidth]{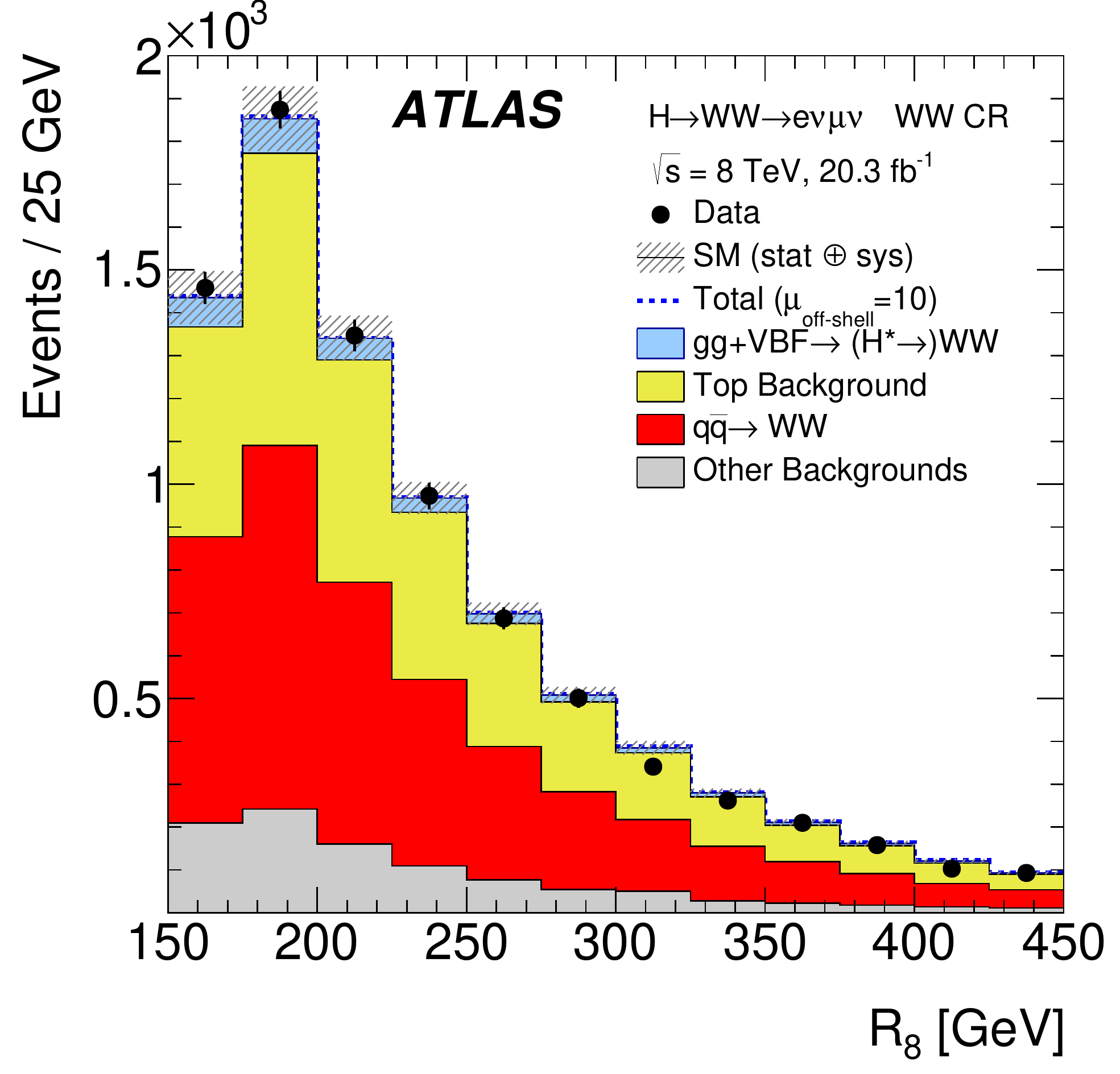}\label{fig:hww_wwcr}}
\subfigure[]{\includegraphics[width=0.55\figwidth]{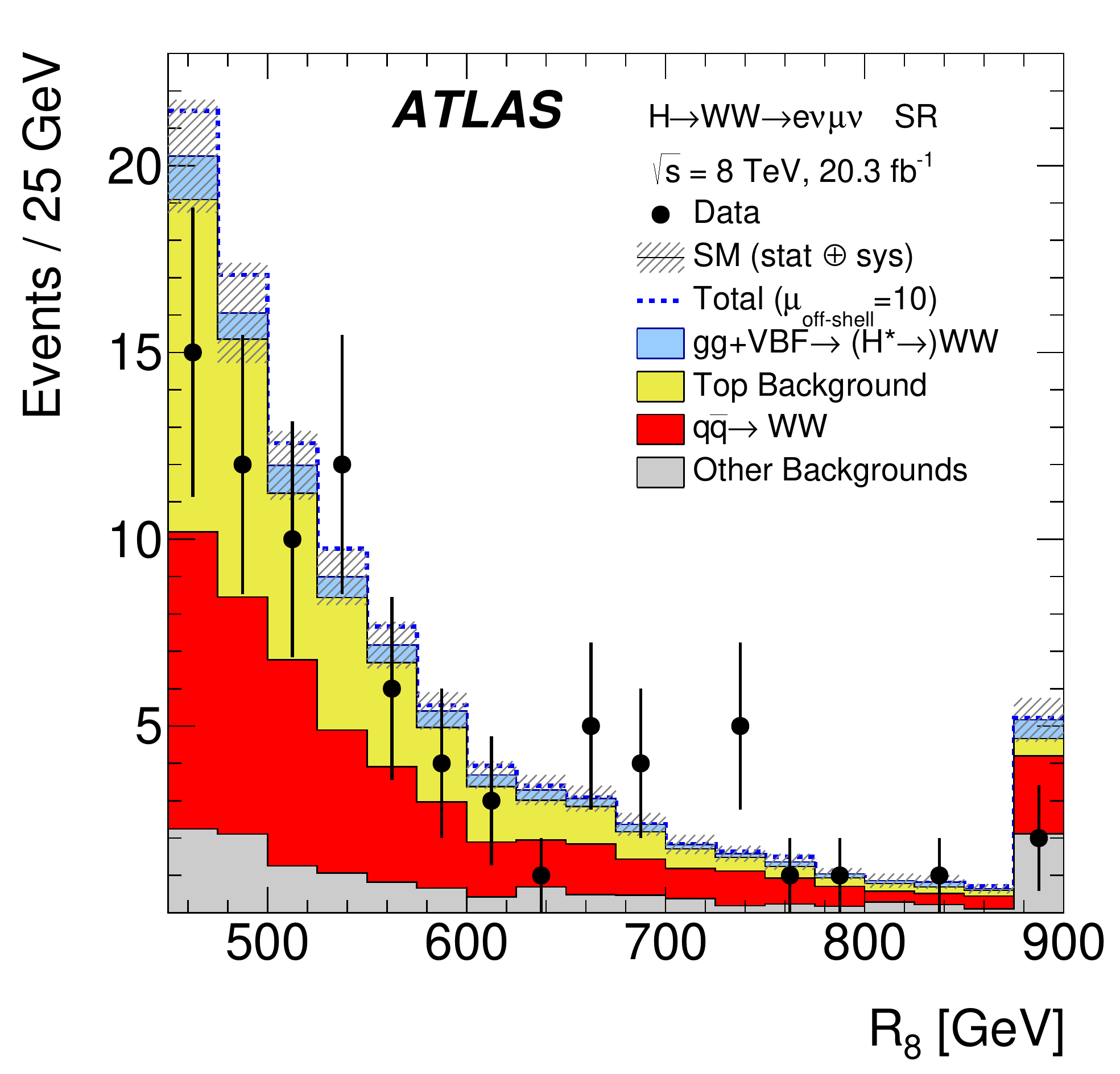}\label{fig:hww_r8}}

\caption{Observed distributions of R$_8$, constructed from the dilepton invariant mass and transverse mass, Eq.~\ref{eq:r8}, in the $WW\rightarrow e\nu\mu\nu$ channel for \subref{fig:hww_topcr} the top control region, \subref{fig:hww_wwcr} $WW$ control 
region (the CRs start at 160~\GeV), and \subref{fig:hww_r8} the signal region for R$_8$ above 450~\GeV,
compared to the expected contributions from the SM including the Higgs boson (stack).
The dashed line corresponds to the total expected event yield, including all backgrounds and the Higgs boson with $\mu_\text{off-shell}=10$.
  The last bin in \subref{fig:hww_topcr} and \subref{fig:hww_r8} includes the overflow.  A relative $gg\rightarrow WW$ background K-factor of R$^B_{H^*}=1$ is assumed.
The top-quark and $WW$ backgrounds are normalised to data as described in Sect.~\ref{sec:hww_event_selection}.  The stacking order follows the legend in each plot.}

\end{figure}

In order to isolate the off-shell Higgs boson production while minimising the impact of higher-order QCD effects on $gg\rightarrow WW$ kinematics, a new variable, R$_8$, is introduced:
\begin{linenomath}
\begin{equation}
  \label{eq:r8}
  \text{R}_8 = \sqrt{m_{\ell\ell}^2 + \left( a \cdot m_\text{T}^{WW}\right)^2}.
\end{equation}
\end{linenomath}
Both the coefficient $a=0.8$ and the requirement R$_8>450$~\GeV\ are optimised for off-shell signal sensitivity while also rejecting on-shell Higgs boson events, which have relatively 
low values of $m_{\ell\ell}$ and $m_{\text{T}}^{WW}$.  
The predicted on-shell signal contamination is \hwwonshell$ (\text{stat.})$ events.  
The MV1 algorithm, 
 at 85\% efficiency, is used to reject $b$-jets with $\pt>20$~\GeV\ and $|\eta|<2.4$ in order to reject backgrounds containing top quarks.  
A more efficient working point for $b$-jet tagging is used compared to the $\zzn$ analysis because of the need to reject a substantially larger top-quark background.  
A requirement on the separation between leptons, $\Delta\eta_{\ell\ell}<1.2$, suppresses $q\bar{q}$-initiated $WW$ production relative to $gg$-initiated production.
The $b$-jet veto and $\Delta\eta_{\ell\ell}$ requirement are found to have a minimal impact on the $WW$-system kinematics and jet multiplicity in the $gg\rightarrow(H^*\rightarrow) WW$ processes.  
Table~\ref{tab:yield_cut_data} contains the predicted and observed event yields in the signal region, $90\pm4$ and $82$ respectively, in agreement with the SM with a small deficit in data.  
The distribution of the R$_8$ variable in the signal region is shown in Fig.~\ref{fig:hww_r8} for the SM expectation and for a Higgs boson with $\mu_{\text{off-shell}}=10$.

\subsection{Background estimation}
\label{sec:wwBkgEst}

The dominant backgrounds arise from processes with real $W$ bosons in the final state. The two backgrounds with the largest expected event yield are top-quark and $q\bar{q}\rightarrow WW$ production.  Dedicated control regions (CRs) are constructed to normalise these two backgrounds in the signal region with a simultaneous fit.  Uncertainties on the extrapolation from the CRs to the signal region are described in Sect.~\ref{sec:qqvvsyst} and \ref{sec:topsys}.

The top-quark background predictions in the signal and $WW$ control region are both normalised from the same top CR.  A sample of top-quark events is obtained by starting from the signal region and reversing the $b$-jet veto by requiring exactly one $b$-tagged jet.  
This is closer in phase space to the $b$-jet-vetoed signal region than requiring at least one $b$-tag and results in a smaller uncertainty.  The statistical error on the top-quark background normalisation is reduced by expanding the top CR down to R$_8>160$~\GeV\ and dropping the $\Delta\eta_{\ell\ell}$ requirement.  
The impact of these changes is discussed in Sect.~\ref{sec:topsys}.  An event yield of \hwwTcrobs events is observed in the top CR (Fig.~\ref{fig:hww_topcr}), resulting in a fit normalisation factor of \hwwTopcrNF, where the uncertainty includes all systematic sources, including extrapolation uncertainties described in Sect.~\ref{sec:topsys}.  The top CR is approximately \hwwTcrpur pure in top-quark events.

The $q\bar{q}\rightarrow WW$ background is normalised to data using an additional CR.  
The region 160~\GeV~$<\text{R}_8<450$~\GeV\ without the $\Delta\eta_{\ell\ell}$ requirement is used because it has a large $WW$ contribution with negligible on-shell Higgs boson contamination and is adjacent to the signal region.  A $b$-jet veto is applied to reject part of the substantial top-quark contamination.  
An event yield of \hwwWWcrobs events is observed in the $WW$ CR (Fig.~\ref{fig:hww_wwcr}), resulting in a fit normalisation factor of \hwwWWcrNF, including all of the uncertainties as above.  
This CR is approximately \hwwWWcrpur pure in $q\bar{q}\rightarrow WW$, while the leading background of top-quark events contributes \hwwWWcrt.  
The $gg$-initiated $WW$ background is estimated from MC simulation, as discussed in Sect.~\ref{sec:gg2vvsim}.

The remaining background predictions, except for $W$+jets and multi-jet production, are taken from MC simulation, as described in Ref.~\cite{ATLAS:2014aga}.  
The predicted fraction of the total background in the signal region arising from $gg\rightarrow WW$, $W$+jets, and $W\gamma/W\gamma^*/WZ/ZZ$ events is approximately 4\% each, while for $Z$+jets it is 2\%.  
The $W$+jets and multi-jet backgrounds are estimated by applying a data-driven extrapolation factor to CRs with lepton candidates failing the nominal lepton identification and isolation, while passing a loosened requirement \cite{ATLAS:2014aga}.

%-------------------------------------------------------------------------------
\section{Systematic uncertainties}
\label{sec:systematics}
The largest systematic uncertainties for this analysis arise from theoretical uncertainties
on the $gg \to H^* \to VV$ signal process, the $gg/q{\bar q} \to VV$ background processes and the
interference between the $gg\to VV$ signal and background processes.
The electroweak $(H^*\to VV)$ processes in association with two jets
contribute about 10--30\% of the total signal.
The associated theoretical uncertainties due to the missing higher-order corrections and PDF variations
are small for $VH$-like and VBF-like processes $pp\to ZZ + 2j$, and are therefore not included in the analysis.
Compared to the theoretical uncertainties, the experimental uncertainties are small
in the $\zzn$ and $\ww$ analyses and close to negligible in the $ZZ \to 4\ell$ analysis. 
In the $\zzn$ and $\ww$ analyses, uncertainties on the extrapolations from the control regions 
to the signal regions are included. 

\subsection{Systematic uncertainties on $gg \to (H^* \to) VV$}
\label{sec:ggHVVsyst}

The uncertainty from missing higher-order QCD and EW corrections to the off-shell $gg \to H^* \to VV$ signal is estimated
in Ref.~\cite{Passarino:2013bha} as a function of the Higgs boson virtuality, $m_{VV}$, and adopted for this analysis.
The uncertainty is 20--30\% for the high-mass region used in this analysis.
The PDF uncertainty for the $gg\to (H^* \to) VV$ process as a function of $m_{VV}$ 
is found to be 10--20\% in the high-mass region used in this analysis. 
This is consistent with an earlier study at $\sqrt{s}=7~\TeV$~\cite{LHCHiggsCrossSectionWorkingGroup:2012vm}. 

For the $gg\to VV$ background, higher-order QCD calculations are not available.
As discussed in Sect.~\ref{sec:gg2vvsim}, the gluon-induced part of the signal K-factor \KHggmVV\ is applied to the background and 
results are then given as a function of the unknown K-factor ratio \Kratio\ between background and signal. 
The uncertainty on \KHggmVV\ is larger than the uncertainty on \KHmVV 
because some contributions to the full signal NNLO QCD K-factor are not present in \KHggmVV. 
Therefore, the following correlation treatment of uncertainties is applied: 
the uncertainty on the signal K-factor \KHmVV\ is applied as a correlated uncertainty to \KHggmVV. 
The difference in quadrature between the uncertainty on \KHggmVV\ and \KHmVV\ is added as an uncorrelated uncertainty component only to \KHggmVV.

The interference between $gg\to H^*\to VV$ and $gg\to VV$ is calculated at LO only.
In Ref.~\cite{Bonvini:2013jha}, a soft-collinear approximation is used to
calculate the cross-section for the sum of a heavy Higgs boson ($gg\to H\to WW$) and its interference with the background.
The uncertainty on this calculation is estimated to be about 10\%,
which leads to about $30\%$ uncertainty on the interference alone. 
Within the ansatz of using an unknown K-factor ratio between background and signal 
(see Eq.~(\ref{eqn:sigma_ggHVVscaling_on_S_B_I})), 
this additional uncertainty of roughly 30\% on the interference term can be represented by
an approximately 60\% variation of the K-factor ratio \Kratio\ for the background around the nominal value of 1.0.
Therefore the variation of \Kratio\ from 0.5 to 2.0 should cover both the leading corrections and uncertainties for the 
interference and the background component taken individually.

However, there is a large cancellation between the background and the
negative interference at the expected 95\% confidence level upper limit value of $\muoffshell$,
shown in Tables~\ref{tab:obs_ul_cls_nominal} and \ref{tab:limitOFFshell_ZZWW}.
This leads to a large artificial cancellation in the uncertainties of the
$gg\to ZZ$ background and the interference, when treated as correlated.
To account for additional uncertainties on the interference component
that are not covered by the soft-collinear approximation,
the 30\% uncertainty on the interference derived in Ref.~\cite{Bonvini:2013jha}
is applied to the interference component in addition to, and uncorrelated with, other uncertainties.

The systematic uncertainties associated with SHERPA-based re-weighting in
$\pt$ of the $VV$ system are assessed by
varying the renormalisation, factorisation and resummation scales in SHERPA.
The larger in value between the scale variations in SHERPA and
50\% of the difference between SHERPA and gg2VV+ PYTHIA8 is assigned as the
systematic uncertainty.
This conservative approach is chosen to consider potential uncertainties not accounted for by the scale variations.
The impact of the PDF uncertainties is found to be negligible.

\subsection{Systematic uncertainties on $q{\bar q} \to VV$}
\label{sec:qqvvsyst}

The missing-higher-order and PDF uncertainties for the $q{\bar q} \to ZZ$ background, 
as a function of $m_{ZZ}$, are taken from Ref.~\cite{LHCHiggsCrossSectionWorkingGroup:2012vm}, 
based on NLO 7~TeV calculations using a fixed scale of $m_Z$. 
Slightly smaller systematic uncertainties are found for 8~\TeV\ using a dynamic scale of $m_{ZZ}/2$, 
hence applying the uncertainties from Ref.~\cite{LHCHiggsCrossSectionWorkingGroup:2012vm} can be considered a conservative choice. 
Both the QCD scale uncertainty and the PDF uncertainty are 5--10\% for the high-mass region used in this analysis.
The NNLO calculation in Ref.~\cite{Cascioli:2014yka} does not yield a significantly reduced QCD scale systematic uncertainty. 
An evaluation of the PDF uncertainty correlations shows that the $q{\bar q} \to ZZ$ background PDF uncertainties are 
anti-correlated with the PDF uncertainties for the $gg \to (H^* \to) ZZ$ process, and this is taken into account in the analysis.
Acceptance uncertainties on the $q{\bar q} \to ZZ$ background are evaluated by comparing PYTHIA8 and HERWIG6~\cite{Corcella:2000bw} samples 
and found to be negligible.
The PDF, QCD scale, and EW correction uncertainties for the $q{\bar q}\to WZ$ process
are considered in the same way as for the $q{\bar q}\to ZZ$ process. 
Both the QCD scale uncertainty and the PDF uncertainty are estimated to be 
$\sim$5--10\% for the high-mass region used in this analysis.

Extrapolation uncertainties on the $q\bar{q} \to WW$ process in the $\ww$ channel are
evaluated using the method described in Ref.~\cite{ATLAS:2014aga}.  Uncertainties due to missing higher-order corrections are
estimated by varying the renormalisation and factorisation scales independently by factors of one-half and two,
keeping the ratio of the scales between one-half and two.  Parton shower and matrix-element uncertainties are
estimated by comparing POWHEG-BOX+PYTHIA8 with POWHEG-BOX+HERWIG6 and POWHEG-BOX+HERWIG6 with aMC@NLO~\cite{Alwall:2014hca}+HERWIG6, respectively. 
PDF uncertainties are estimated by taking the largest difference between the nominal CT10~\cite{Lai:2010vv} and either 
the MSTW2008~\cite{Martin:2009iq} or the 
NNPDF2.1~\cite{Ball:2011mu} PDF set and adding this in quadrature with the CT10 error eigenvectors (following the procedure described in Ref.~\cite{Botje:2011sn}). 
The extrapolation uncertainties from the $WW$ control region to the signal region are summarised in Table~\ref{tab:hww_sys}.

\begin{table}[!htbp]
\centering
  \begin{tabular}{|l|cccc|}
   \hline
     & UE/PS & Gen. & Scale & PDF \\
    \hline
    Top CR&6.4\% & 2.4\% & 2.4\% & 2.4\% \\
    $WW$ CR & 2.5\% & 2.8\% & 2.3\% & 1.5\% \\
    \hline
  \end{tabular}
\caption{Uncertainties on the extrapolation of top-quark processes and $q\bar{q}\rightarrow WW$ from their respective CRs to the SR, and from the Top CR to the $WW$ CR, from the parton shower and underlying event (UE/PS), from matching the matrix element to the UE/PS model (Gen), from the QCD renormalisation and factorisation scale (Scale), and from the PDFs.  These uncertainties are used in the $WW$ analysis and derived with the same methods as used in Ref.~\cite{ATLAS:2014aga}.}
  \label{tab:hww_sys}
\end{table}

The EW corrections for the $q\bar{q} \to VV$ process described in Sect.~\ref{sec:qqVV} 
are strictly valid only for the LO QCD $q\bar{q} \to VV$ process above the
diboson production threshold when both vector bosons are on shell.   
This is the case for all three analyses after final selections. 
The EW corrections are computed at LO QCD because the mixed QCD--EW corrections have not yet been calculated.
In events with high QCD activity, an additional systematic uncertainty is considered by 
studying the variable $\rho=\left|\sum_i\vec{\ell}_{i,\text{T}}+\vec{E}_{\mathrm{T}}^{\mathrm{miss}}\right|/\left(\sum_i\left|\vec{\ell}_{i,\text{T}}\right|+\left|\vec{E}_{\mathrm{T}}^{\mathrm{miss}}\right|\right)$
introduced in Eq.~(4.4) of Ref.~\cite{Gieseke:2014gka} 
(here $\vec{\ell}_{T}$ represents the transverse momentum of the lepton $i$ from vector boson decays). 
A phase space region with $\rho<0.3$ is selected, where the NLO QCD event kinematics resembles 
the LO event kinematics in being
dominated by recoiling vector bosons and therefore the corrections are applicable without additional uncertainty. 
For events with $\rho>0.3$ the correction is applied with a 100\% systematic uncertainty
to account for the missing mixed QCD--EW corrections which are expected to be of the same order of magnitude.
The applied corrections are partial in that they include only virtual corrections, 
and do not include polarisation effects. 
The sum of both of these effects is estimated to be $\mathcal{O}(1\%)$~\cite{Gieseke:2014gka} and is neglected in this analysis.

While the EW corrections and uncertainties directly affect the predicted size of the 
$q\bar{q} \to ZZ$ and $q\bar{q} \to WZ$ backgrounds in the $ZZ \to 4\ell$ and $\zzn$ analyses, 
only the extrapolation of the $q\bar{q} \to WW$ background from the control region to the signal region is affected
in the $\ww$ analysis.

\subsection{Systematic uncertainties on top-quark events}
\label{sec:topsys}

Theory uncertainties on extrapolating top-quark processes from the control region to the signal region 
in the $\ww$ channel are also evaluated using methods similar to those of Ref.~\cite{ATLAS:2014aga}.
For the evaluation of the extrapolation uncertainties, the signal region requirements are relaxed to increase the sample size; the region is extended down to R$_8>160$~GeV and the $\Delta\eta_{\ell\ell}$ requirement is dropped.
The extra uncertainty from this extension is checked in a separate sample with at least one $b$-tagged jet,
again defined so as to reduce the statistical uncertainties, which is simultaneously re-weighted in $\Delta\eta_{\ell\ell}$ and R$_8$ to match the $b$-vetoed region.
With this $b$-tagged sample, the extra uncertainty from the removal of the $\Delta\eta_{\ell\ell}$ requirement, and from extending the range in R$_8$, is found to be 3.5\%.

The method described in Sect.~\ref{sec:qqvvsyst} is used to evaluate
the systematic uncertainties on top-quark processes.
Since the extended signal region covers the $WW$ CR,
the same systematic uncertainties are valid for the extrapolation from the top CR to the $WW$ CR.
These uncertainties, summarised in Table~\ref{tab:hww_sys}, are applied to both $t\bar{t}$ and 
single-top processes,
which make up approximately \hwwstfrac of the top background in the signal region.
A \hwwstsys  uncertainty is assigned to the single-top processes in order to take into account the uncertainty on the single-top fraction; the impact on the result is negligible.

\subsection{Experimental systematic uncertainties}
For the $ZZ\to4\ell$ analysis, the same sources of experimental uncertainty as in Ref.~\cite{Aad:2014eva} are evaluated.
In the off-shell Higgs boson region, the leptons come from the decay of on-shell $Z$ bosons;
hence the lepton-related systematic uncertainties are small compared
to those for the leptons from on-shell Higgs boson production.
The leading, but still very small, experimental systematic uncertainties are due to the electron and muon
reconstruction efficiency uncertainties.

Similarly, for the $2\ell 2\nu$ channel, the same sources of experimental uncertainty as in Ref.~\cite{Aad:2014iia} are evaluated.
The electron energy scale, electron identification efficiency,
muon reconstruction efficiency, jet energy scale, and systematic uncertainties from the data-driven $Z$ background
estimates are the main sources of the experimental systematic uncertainties.
These experimental uncertainties affect the expected sensitivity of the
\muoffshell\ measurement only at the percent level. 

Finally, for the $\ww$ channel, the same sources of experimental uncertainty as in Ref.~\cite{ATLAS:2014aga} are evaluated.  The uncertainty on the electron energy scale, followed by the uncertainty on the rate for mis-tagged light-flavour jets as $b$-jets, and the uncertainty on the jet energy scale and resolution, are the dominant experimental sources of uncertainty.  The remaining experimental sources are significantly smaller than the theoretical uncertainties.  

The uncertainty on the integrated luminosity is $2.8\%$. It is derived, following the same methodology as 
that detailed in Ref.~\cite{lumi2011}, from a preliminary calibration of the luminosity 
scale derived from beam-separation scans performed in November 2012.

%-------------------------------------------------------------------------------
\section{Results}
\label{sec:results}
In this section the results for the $ZZ \to 4\ell$, $\zzn$ and  $\ww$ analyses are presented
and translated into limits on the off-shell signal strength $\mu_\text{off-shell}$ for the individual analyses
and for the combination of all three channels.
In a second step, the off-shell analyses are combined with the on-shell $ZZ^{*} \to 4 \ell$~\cite{Aad:2014eva} 
and $WW^{*} \to  \ell \nu \ell \nu$~\cite{ATLAS:2014aga} analyses based on the 8~\TeV\ data taken in 2012.
In combining the $ZZ$ and $WW$ channels it is assumed that the ratio 
of the $ZZ$ cross-section $\sigma^{gg \to H^{(*)} \to ZZ}(\hat{s})$ to the $WW$ cross-section
$\sigma^{gg \to H^{(*)} \to WW}(\hat{s})$ (and similarly for VBF) is as predicted in the SM for both the on- and off-shell processes.
  
Two different off-shell combinations are presented based on different assumptions. 
First, a single off-shell signal strength parameter is applied for all production modes. 
This is equivalent to assuming that the ratio of 
the off-shell production rates via the  process $gg \to H$ to those via the VBF process are as predicted in the SM.
In a second combination, only the off-shell signal strength of the $gg \to H^* \to VV$ production process is considered 
while the VBF production process is fixed to the SM prediction. 
In this case the combined signal strength $\muoffshell^{gg\to H^* \to VV}$ can be interpreted as a constraint on the 
off-shell coupling strength $\kappa_{g,\text{off-shell}}$ associated with the $gg\to H^*$ production mode. 

The combination with the on-shell analyses is also performed under two assumptions 
that correspond to different interpretations of the results. 
The first is performed using different signal strengths for the $gg \to H^{(*)}$ and the VBF production modes.%
\footnote{In all results the signal strength for $VH$ associated production is assumed 
to scale with VBF production while the $b\bar{b}H$ and $t\bar{t}H$ processes scale with the $gg\to H$ process. 
These additional production modes are expected to give negligible contributions to 
the off-shell measurements, but have small contributions to the on-shell signal yields.}
The parameter of interest is described by the ratio of the off-shell to the on-shell signal strengths, 
which can be interpreted as the Higgs boson total width normalised to its SM prediction: $\muoffshell/\muonshell=\Widthratio$.
This interpretation requires that the off- and on-shell couplings are the same for both $gg\to H^{(*)}$ and VBF production modes 
(i.e., $\kappa_{g,\text{on-shell}}=\kappa_{g,\text{off-shell}}$ and $\kappa_{V,\text{on-shell}}=\kappa_{V,\text{off-shell}}$~\footnote{To set an upper 
limit, the assumption in Eq.~(\ref{eqn:kappa_onshell<offshell}), and the equivalent assumption for the VBF production mode, is sufficient.}).
In a second combination, the coupling scale factors $\kappa_{V}=\kappa_{V,\text{on-shell}}=\kappa_{V,\text{off-shell}}$ 
associated with the on- and off-shell VBF production and the $H^{(*)} \to VV$ decay, 
are assumed to be the same and fitted to the data (profiled).
In this case the parameter of interest, $R_{gg}=\muoffshell^{gg\to H^*}/\muonshell^{gg \to H}$, can be interpreted as the 
ratio of the off-shell to the on-shell gluon couplings: $R_{gg}=\kappa^2_{g,\text{off-shell}}/\kappa^2_{g,\text{on-shell}}$. 
This also assumes that the total width is equal to the SM prediction. 

In the $ZZ\to 4\ell$ channel, a binned maximum-likelihood fit to the ME-based discriminant distribution is performed to
extract the limits on the off-shell Higgs boson signal strength. 
The fit model accounts for signal and background processes, 
including $gg\to (H^*\to) ZZ$, VBF$(H^*\to) ZZ$ and $q{\bar q}\to ZZ$. 
The probability density functions (pdf) of the signal-related processes 
$gg\to (H^*\to) ZZ$ and VBF $(H^*\to) ZZ$ are parameterised as a function of 
both the off-shell Higgs boson signal strength \muoffshell\ and the unknown background K-factor ratio \Kratio\ 
as given in Eqs.~(\ref{eqn:ggHZZscaling_on_S_SBI_B}) and (\ref{eq:pdf_VBFzz}). 
Normalisation and shape systematic uncertainties on the signal and background processes are taken into account 
as described in Sect.~\ref{sec:ggHVVsyst}, 
with correlations between different components and processes as indicated therein.

In the $ZZ\to 2\ell2\nu$ channel, a similar maximum-likelihood fit to the transverse mass ($m_{\mathrm{T}}^{ZZ}$) is performed,
comparing the event yield in the signal-enriched region in data with the predictions.
The fit model accounts for the signal 
and all background processes mentioned in Table~\ref{tab:yield_cut_data}.
The modelling of the dominant signal and background processes is the same as in the $ZZ\to 4\ell$ channel.

In the $\ww$ channel, a maximum-likelihood fit is performed using the event yields in the signal region and the two control regions.  
As in the $ZZ$ channels, the fit model accounts for the parameterised signal and all background processes mentioned in Sect.~\ref{sec:wwBkgEst}.  
Unconstrained strength parameters common among fit regions for the $q\bar{q}\rightarrow WW$ and top-quark processes 
allow the control regions to constrain the predicted event yields in the signal region.  

The likelihood is a
function of a parameter of interest $\mu$ and nuisance parameters $\vec\theta$.  
Hypothesis testing and
confidence intervals are based on the profile likelihood
ratio~\cite{Cowan:2010st}.  The parameters of interest are different in the various tests, 
while the remaining parameters are profiled.
Hypothesised values for a parameter of interest $\mu$ are tested with a
statistic
\begin{linenomath}
\begin{equation}\label{eq:teststatMhProf}
   \Lambda(\mu) = \frac{L\big(\mu,  
\hat{\hat{\vec\theta}}(\mu)\big)} {L(\hat{\mu},
\hat{\vec\theta})} \quad ,
\end{equation}
\end{linenomath}
where the single circumflex denotes the unconditional maximum-likelihood estimate of a parameter and the double circumflex
({\it e.g.} $\hat{\hat{\vec\theta}}(\mu)$) denotes the conditional
maximum-likelihood estimate ({\it e.g.} of $\vec\theta$) for given
fixed values of $\mu$.  This test statistic extracts the
information on the parameters of interest from the full likelihood
function.  

All 95\% confidence level (CL) upper limits are derived using the \CLs\ method~\cite{Read:2002hq}, 
based on the following ratio of one-sided $p$-values:
$\CLs(\mu) = p_{\mu}/({1-p_{1}})$
where $p_{\mu}$ is the $p$-value for testing a given $\mu=\muoffshell$ or
$\mu=\Widthratio$ (the non-SM hypothesis) and $p_{1}$ is the $p$-value derived
from the same test statistic under the SM hypothesis of $\muoffshell=1$ in the first case and 
$\Widthratio =\muonshell=1$ in the second case.%
\footnote{In the context of this analysis the alternative hypothesis is given by the SM value(s) 
for all relevant parameters of the fit model.}
The 95\% \CLs\ upper limit is found by solving for $\CLs(\mu^\text{95\%})=5\%$. 
Values $\mu > \mu^\text{95\%}$ are regarded as excluded at 95\% CL. 
A detailed description of the implementation of the \CLs\ procedure can be found in Ref.~\cite{paper2012prd}.

The results presented in this paper rely on the asymptotic approximation~\cite{Cowan:2010st} for the test statistic $\Lambda(\mu)$.
This approximation was cross-checked with Monte Carlo ensemble tests that confirm its validity in
the range of the parameters for which the 95\% CL limits are derived. 
Deviations appear close to the boundary of $\muoffshell \ge 0$ imposed by Eq.~(\ref{eqn:sigma_ggHVVscaling_on_S_B_I}) 
and hence the $1\sigma$ uncertainties can only be seen as approximate.

While the final 95\% CL limits are given as a function of the unknown background K-factor ratio \Kratio, 
comparisons between the data and the MC predictions, and values in other figures and tables, 
are given assuming \Kratio=1.

\subsection{Results of the individual off-shell analyses}

\begin{figure}[!htbp]
\begin{center} 
\subfigure[]{\includegraphics[width=0.55\figwidth]{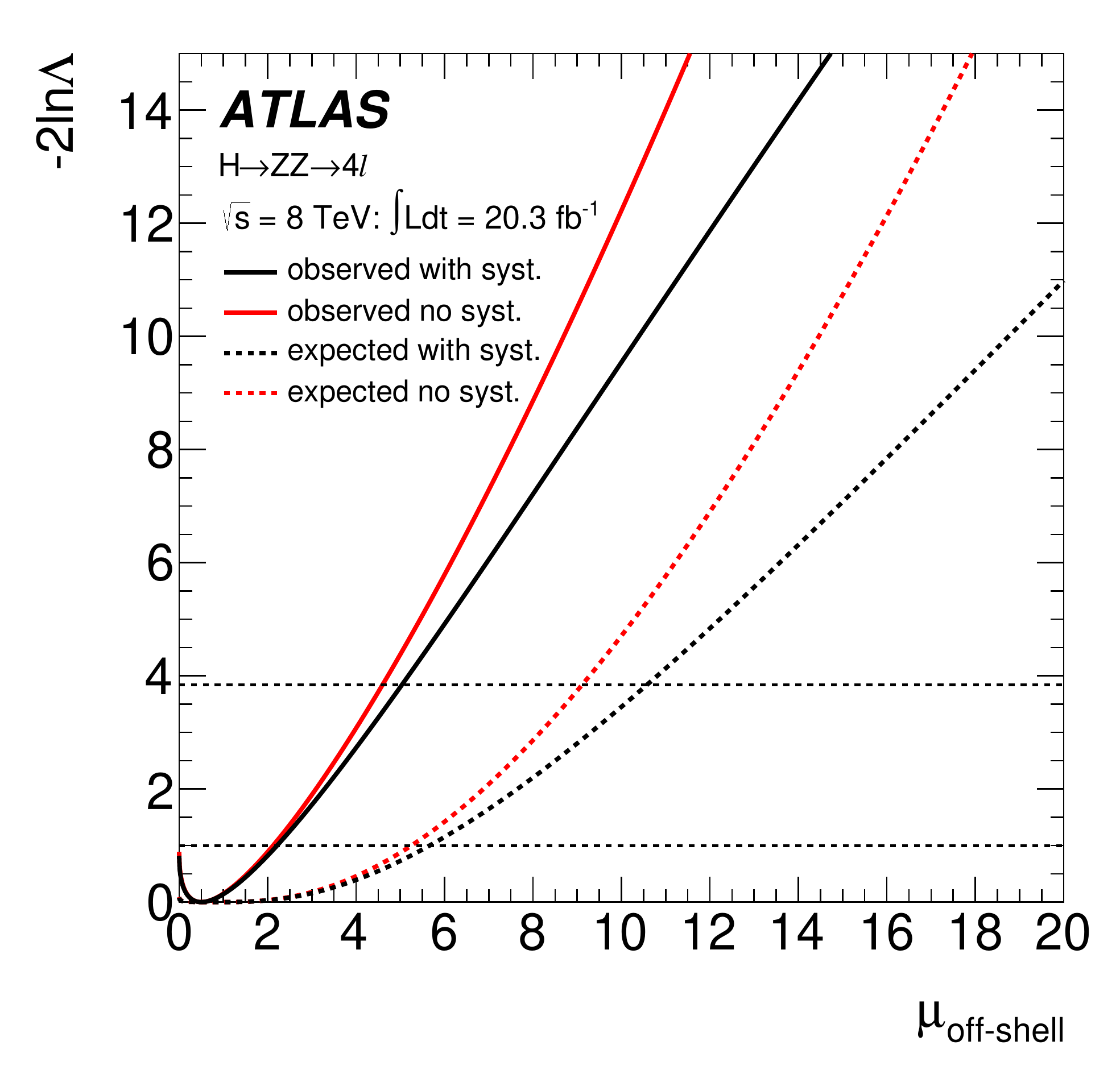}\label{fig:zz4l_results:lnl}}
\subfigure[]{\includegraphics[width=0.55\figwidth]{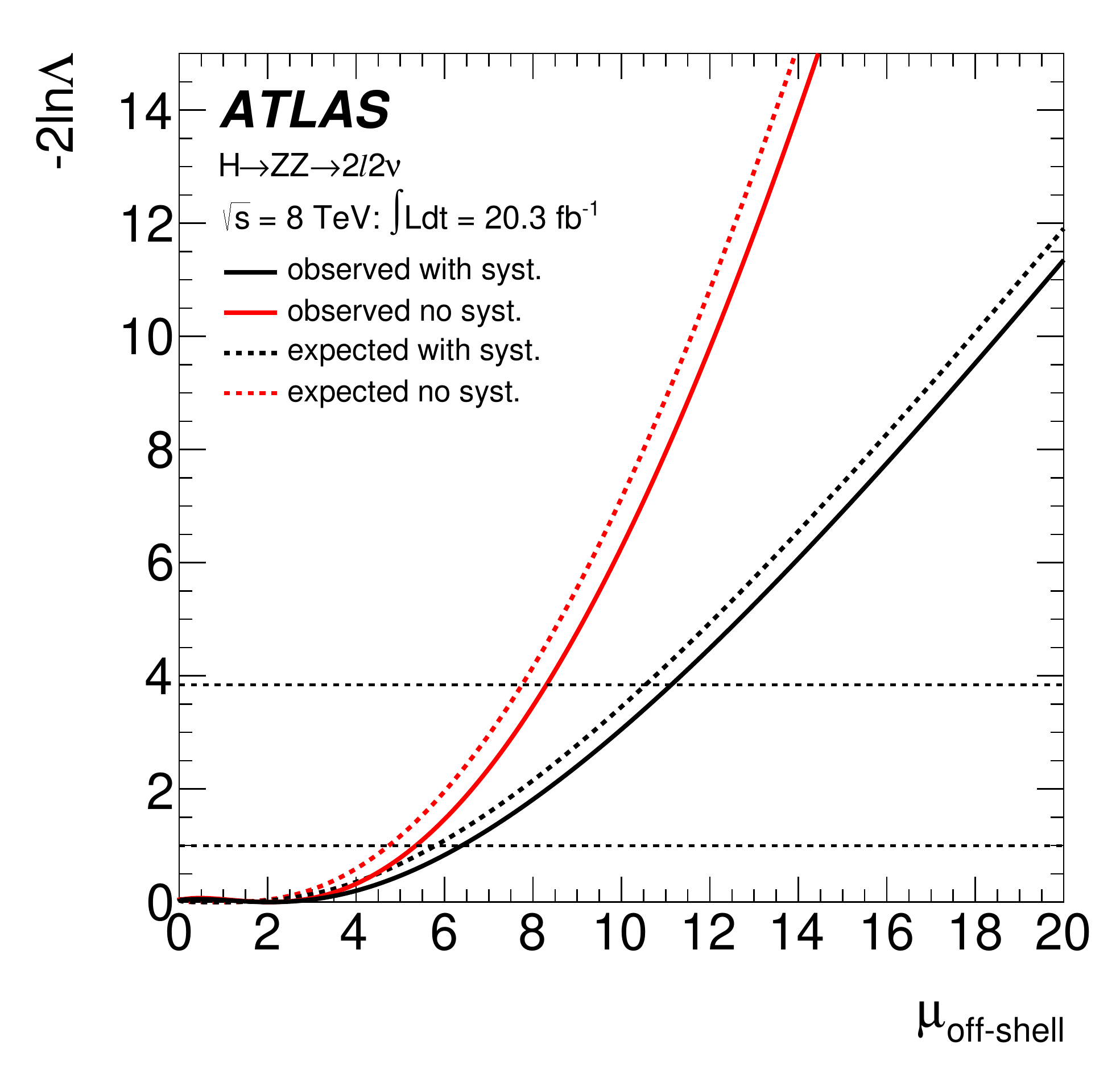}\label{fig:2l2n_results:lnl}} \\
\subfigure[]{\includegraphics[width=0.55\figwidth]{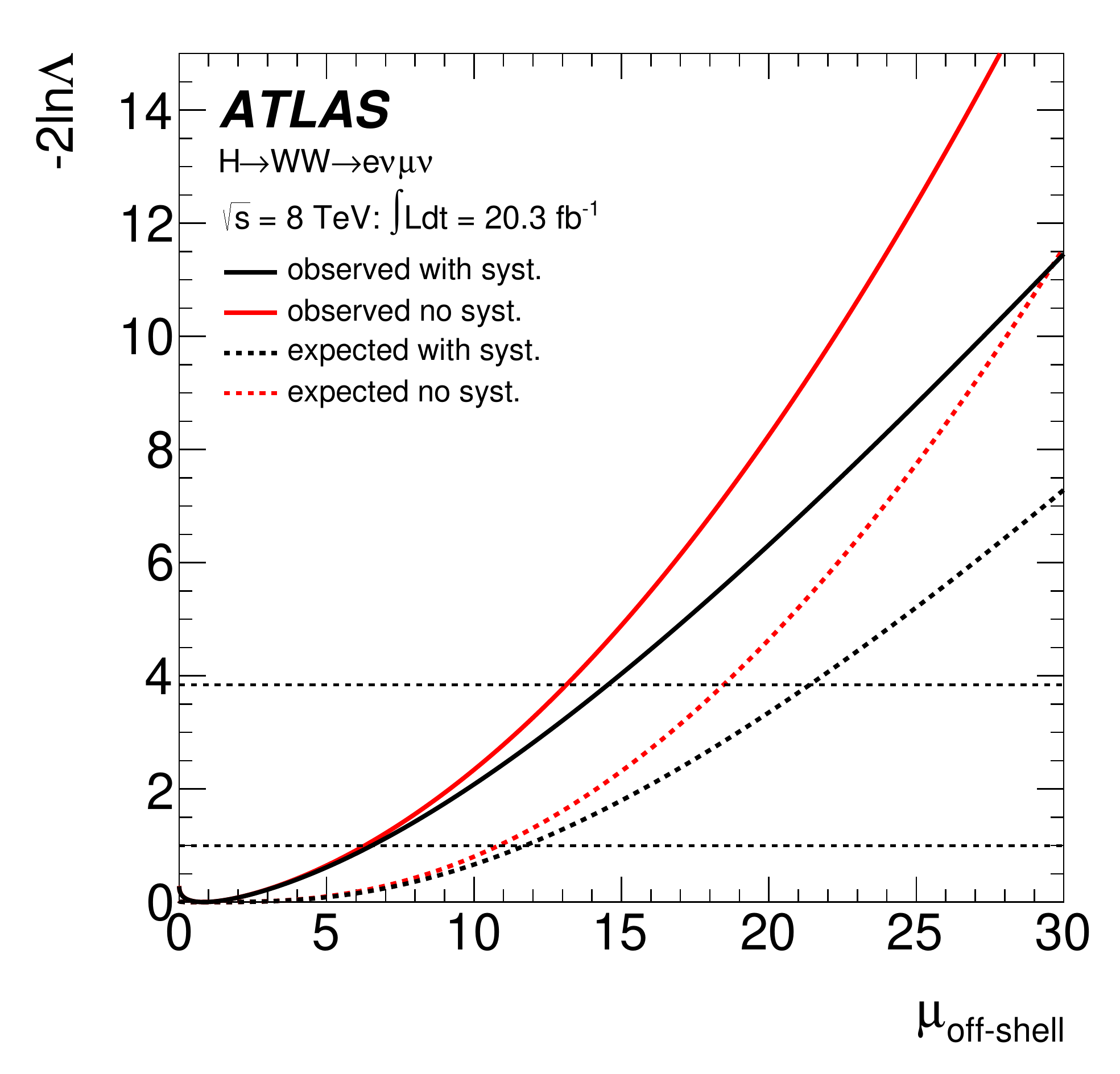}\label{fig:lvlv_results:llh}}
\caption{Scan of the negative log-likelihood, $-2\ln\Lambda$, as a function of $\mu_\text{off-shell}$, 
in the $ZZ\to 4\ell$ (a),  $\zzn$ (b) and $\ww$ (c) channels.
The black solid (dashed) line represents the observed (expected) value including all systematic uncertainties,
while the red solid (dashed) line is for the observed (expected) value without systematic uncertainties.
A relative $gg\to VV$ background K-factor of \Kratio=1 is assumed in these figures.
}
\label{fig:individual_results_lln}
\end{center}
\end{figure}

\begin{figure}[!htbp]
\begin{center}
\subfigure[]{\includegraphics[width=0.55\figwidth]{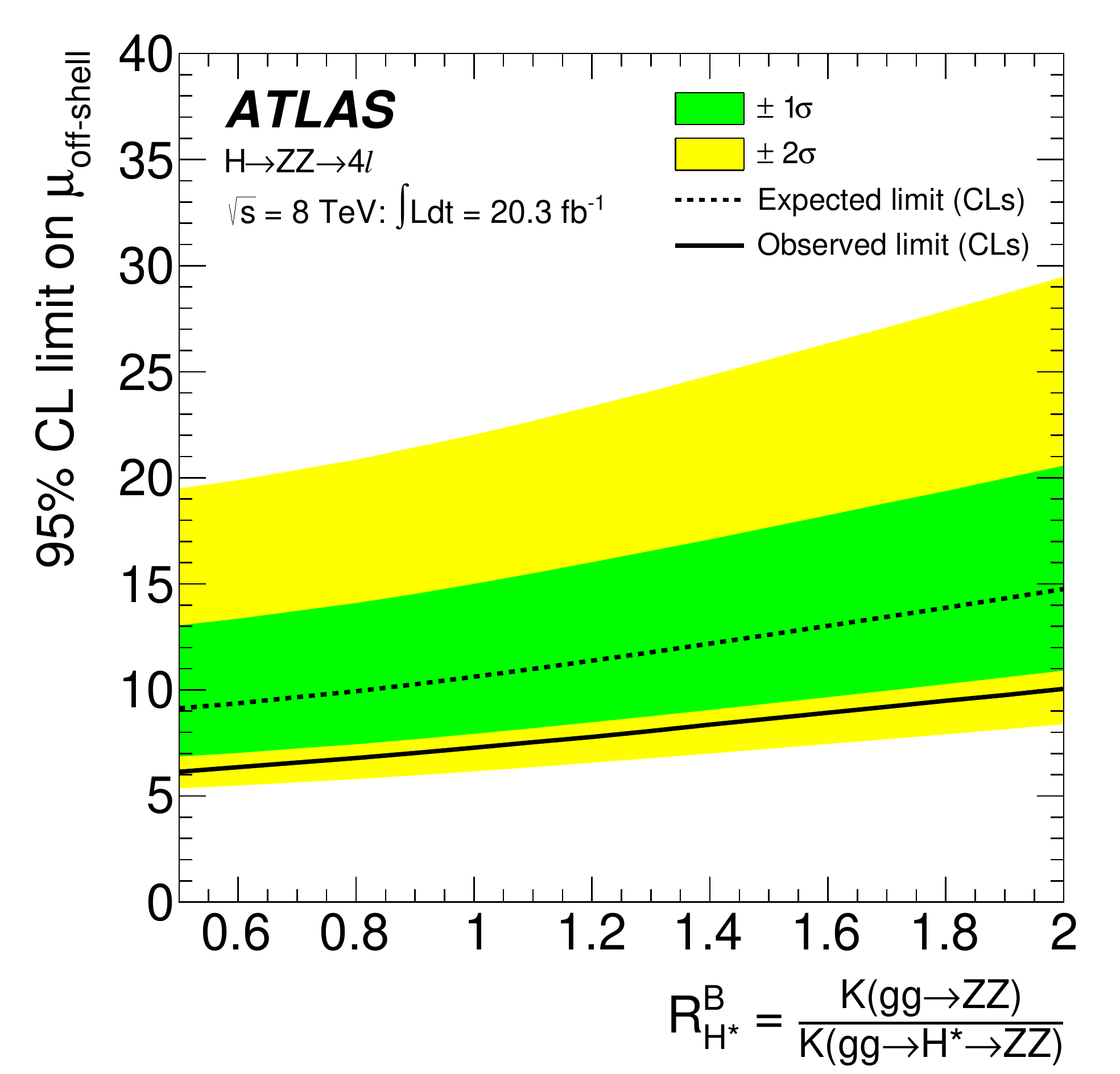}\label{fig:zz4l_results:ul}}
\subfigure[]{\includegraphics[width=0.55\figwidth]{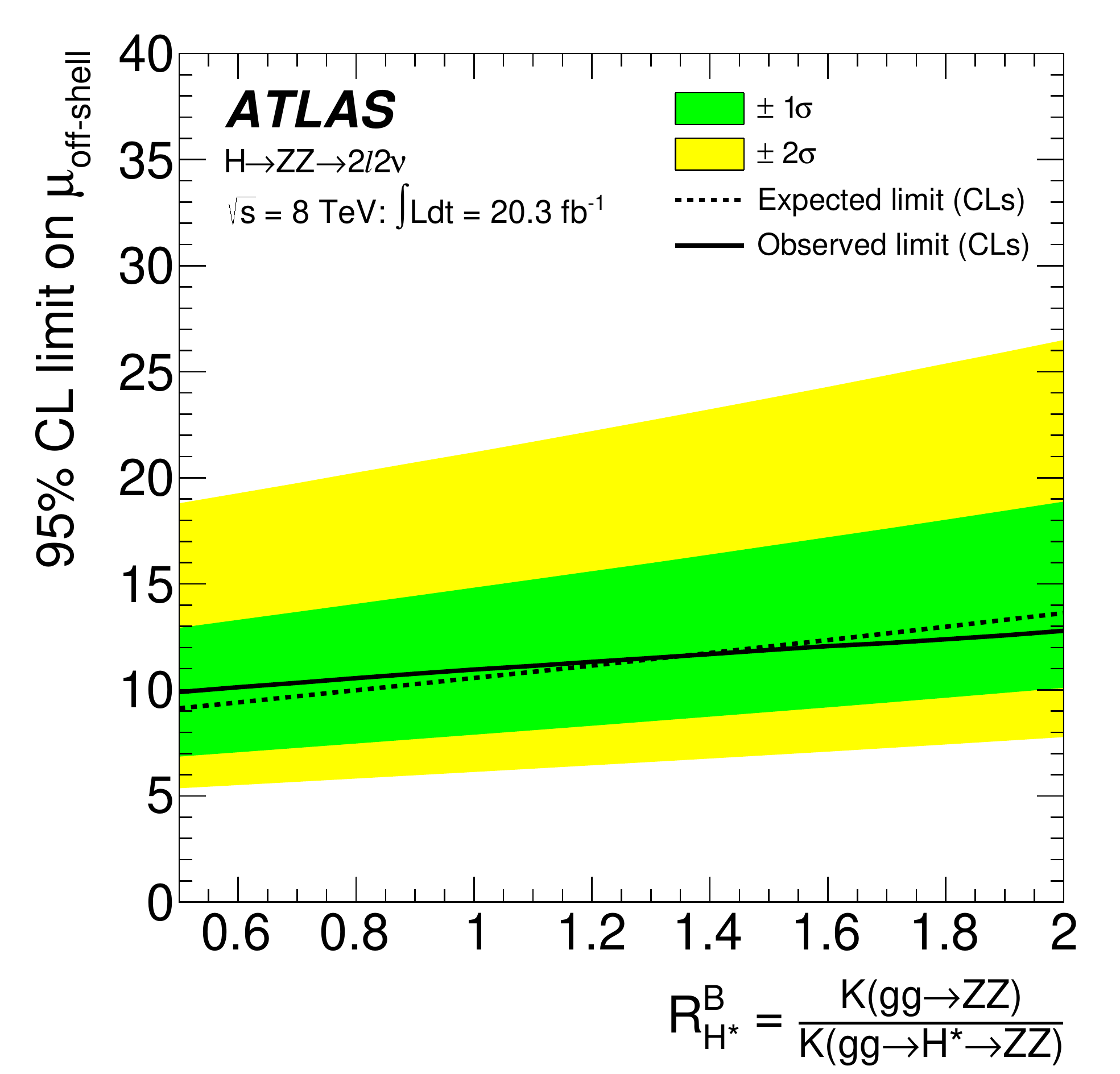}\label{fig:zz2l2n_results:ul}}
\subfigure[]{\includegraphics[width=0.55\figwidth]{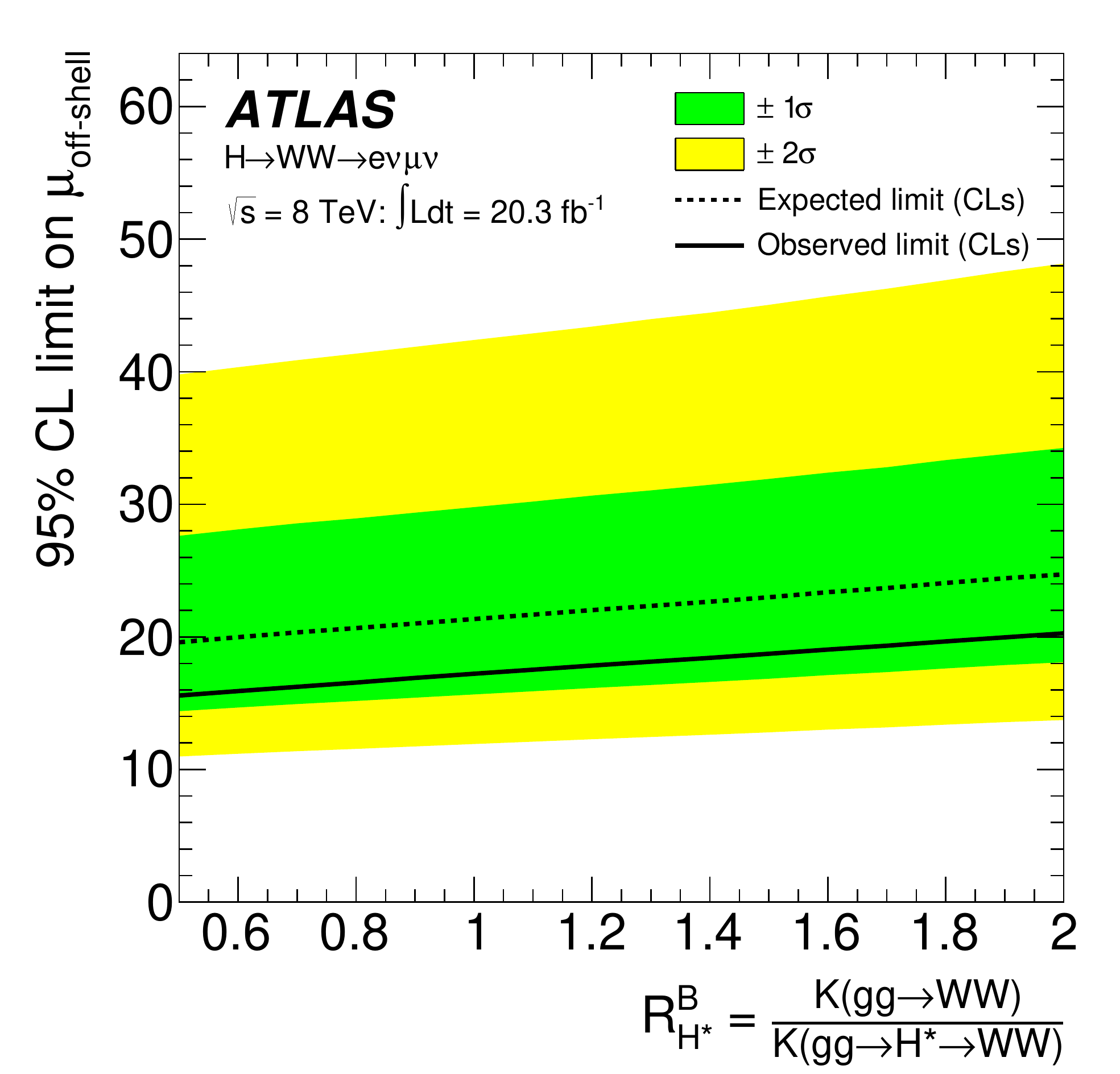}\label{fig:lvlv_results:ul}}
\caption{The observed and expected 95\% CL upper limit on $\mu_\text{off-shell}$ as a function of \Kratio, 
for the $ZZ\to 4\ell$ (a),  $\zzn$ (b) and $\ww$ (c) channels. 
The upper limits  are evaluated using the \CLs\ method, with the alternative hypothesis $\mu_\text{off-shell}=1$.
The green (yellow) bands represent the 68\% (95\%) confidence intervals for the \CLs\ expected limit.
}
\label{fig:individual_results_cls}
\end{center}
\end{figure}

The scan of the negative log-likelihood, $-2\ln \Lambda$, as a function of
$\mu_\text{off-shell}$ for data and the expected curve for an SM Higgs boson for the three individual off-shell 
analyses is illustrated in Fig.~\ref{fig:individual_results_lln}.
The observed and expected 95\% CL upper limits on $\mu_\text{off-shell}$ as a function of \Kratio\
are shown in Fig.~\ref{fig:individual_results_cls} and are summarised in Table~\ref{tab:obs_ul_cls_nominal}.
The  $ZZ\to 4\ell$ and $ZZ\to 2\ell 2\nu$ analysis have a very similar expected sensitivity. 
The $ZZ\to 4\ell$ analysis is statistics limited, while the sensitivity in the $ZZ\to 2\ell 2\nu$ analysis is significantly 
reduced by the theoretical systematic uncertainties as can be seen in Fig.~\ref{fig:individual_results_lln}.
The similar expected \CLs\ limits for the two channels for $\Kratio=0.5$ and 1.0 in Table~\ref{tab:obs_ul_cls_nominal} is a 
coincidence, caused by the different statistical and systematic uncertainty components.

\begin{figure}[!htbp]
\begin{center}
\includegraphics[width=0.6\figwidth]{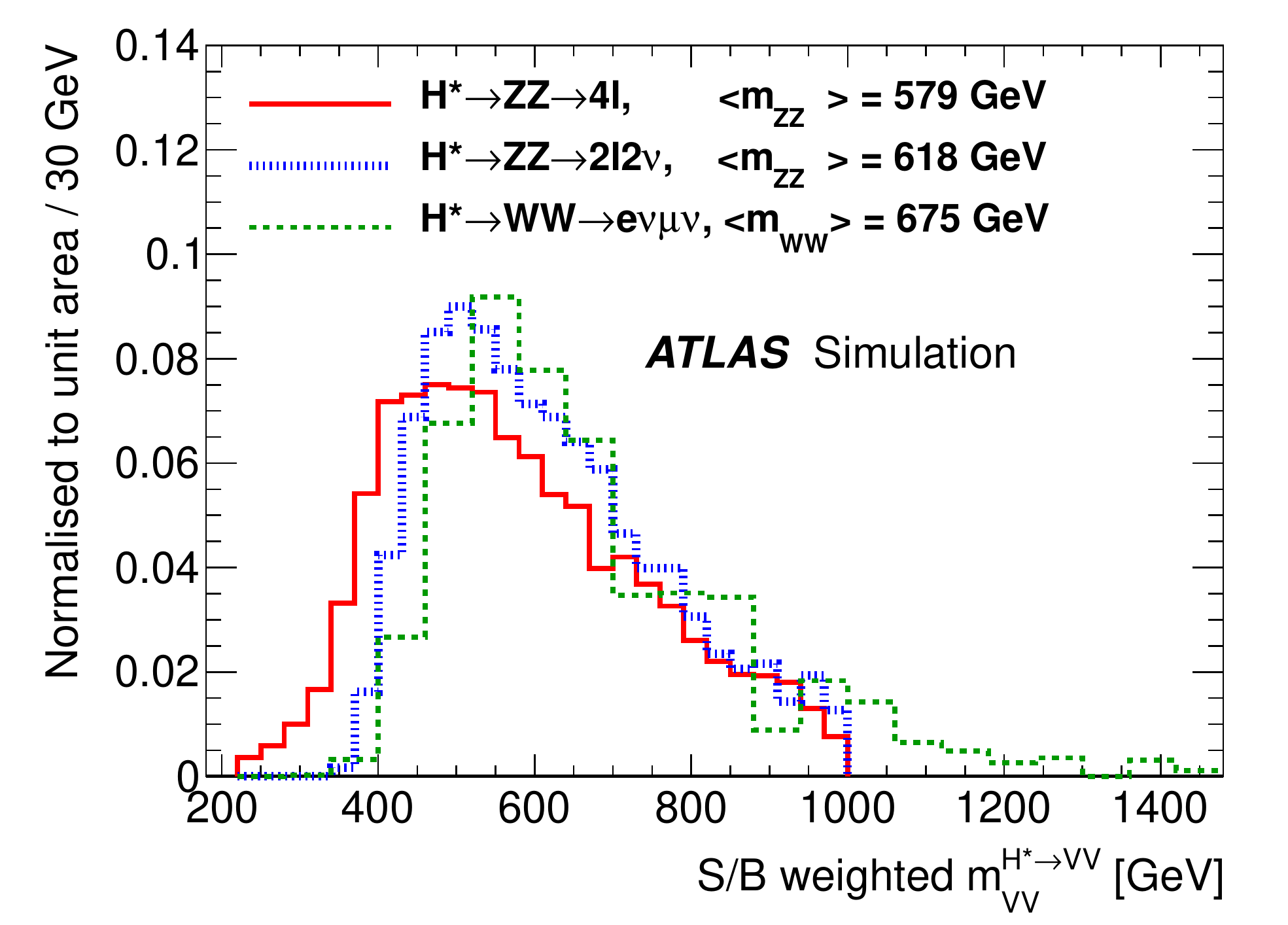}
\caption{Normalised distribution of the generated mass $m_{VV}$ for the $gg \to H^* \to VV$ 
and the VBF $H^* \to VV$ signal processes weighted by the expected S/B ratio in each bin of the final discriminant for the
$ZZ\to 4\ell$ and $ZZ\to 2\ell 2\nu$ analyses and for all events in the signal region for the $\ww$ analysis.
}
\label{fig:truth_mVV_individual_channels}
\end{center}
\end{figure}

The typical off-shell mass scales tested by the analyses are in the range 400~\GeV~$< m_{VV} < 1000~$~\GeV, 
with a small fraction of the expected $H^* \to WW$ signal extending to substantially higher mass scales.%
\footnote{While the $H^* \to ZZ$ analysis includes a selection cut to limit the mass range to $m_{ZZ}\lesssim 1000$~\GeV, 
no such cut can be efficiently implemented for the $H^* \to WW$ analysis due to the poor mass resolution.}
This is illustrated in Fig.~\ref{fig:truth_mVV_individual_channels}, 
which shows the generated $m_{VV}$ mass for the $gg \to H^* \to VV$ 
and the VBF $H^* \to VV$ signal processes weighted by the expected S/B ratio in each bin of the final discriminant for the
$ZZ\to 4\ell$ and $ZZ\to 2\ell 2\nu$ analyses and for all signal events in the signal region for the $\ww$ analysis.

%%%%%%%%%%%%%%%%%%%%%%%%%%%%
\begin{table}[!htbp]
 \centering
 \begin{tabular}{|r| r @{\hspace{4ex}}r @{\hspace{4ex}}r | r @{\hspace{4ex}}r @{\hspace{4ex}}r |}
  \hline
 & \multicolumn{3}{c|}{Observed} & \multicolumn{3}{c|}{Median expected} \\
\Kratio & 0.5 & {\bf 1.0} & 2.0  & 0.5 & {\bf 1.0} & 2.0  \\ [0.5ex]
\hline
\rule[-1.1ex]{0pt}{3.6ex}
$ZZ\to 4\ell$ analysis & 6.1 & {\bf 7.3} & 10.0 & 9.1 & {\bf 10.6} & 14.8 \\
$\zzn$ analysis & 9.9 & {\bf 11.0} & 12.8 & 9.1 & {\bf 10.6} & 13.6 \\
$\ww$ analysis & 15.6 & {\bf 17.2} & 20.3 & 19.6 & {\bf 21.3} & 24.7 \\
\hline
 \end{tabular}
  \caption{The observed and expected 95\% CL upper limits on $\mu_\text{off-shell}$ within the range of $0.5<\Kratio<2.0$.
The bold numbers correspond to the limit assuming $\Kratio=1$.
The upper limits are evaluated using the \CLs\ method, with the alternative hypothesis $\mu_\text{off-shell}=1$.
}
 \label{tab:obs_ul_cls_nominal}
\end{table}
%%%%%%%%%%%%%%%%%%%%%%%%%%

%%%%%%%%%%%%%%%%%%%%%%%%%%
%%%%%%%%%%%%%%%%%%%%%%%%%%

%%%%%%%%%%%%%%%%%%%%%%%%%%
%%%%%%%%%%%%%%%%%%%%%%%%%%

%%%%%%%%%%%%%%%%%%%%%%%%%%
%%%%%%%%%%%%%%%%%%%%%%%%%%

\subsection{Combination of the off-shell $ZZ$ and $WW$ analyses}
The analyses described in the previous sections are combined to obtain a limit on \muoffshell.
In combining the off-shell results the main systematic uncertainties related to the theory 
uncertainties on the $gg \to (H^* \to) VV$ (including signal and interference contributions) 
and $q\bar{q} \to VV$ processes are treated as correlated between the different channels. 
The same K-factor ratio \Kratio\ is assumed for the $gg \to ZZ$ and $gg \to WW$ backgrounds.
Where appropriate, the experimental systematic uncertainties are also treated as correlated. 
However, they are found to have a very small impact on the final combined limit.

The limits on \muoffshell\ are obtained under two different assumptions: 
\begin{itemize}
\item Determination of the signal strength $\mu_{\text{off-shell}}$ when fixing the ratio of the signal strength in $gg \to H^*$ and VBF to the SM prediction, namely $\mu_{\text{off-shell}}^{gg \to H^*} / \mu_{\text{off-shell}}^{VBF}$=1.
\item Determination of the signal strength $\mu_{\text{off-shell}}^{gg \to H^* \to VV}$ when fixing the VBF off-shell signal strength to the SM prediction, i.e. $\mu_{\text{off-shell}}^{\text{VBF }H^* \to VV}$=1.
\end{itemize}

\begin{figure}[!htbp]
\begin{center}
\subfigure[]{\includegraphics[width=0.48\figwidth]{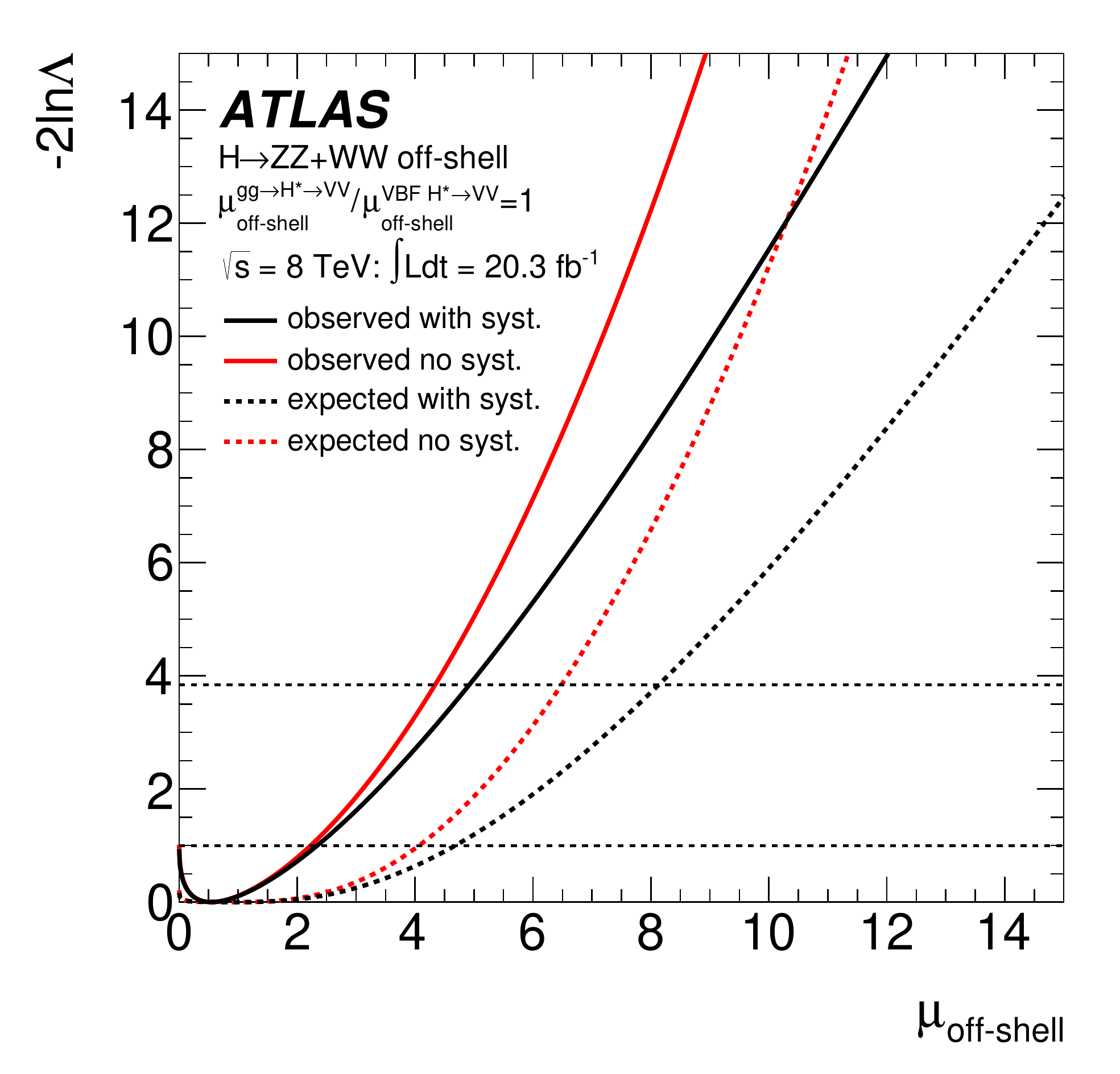}\label{fig:scanRbVBFfree}}
\subfigure[]{\includegraphics[width=0.48\figwidth]{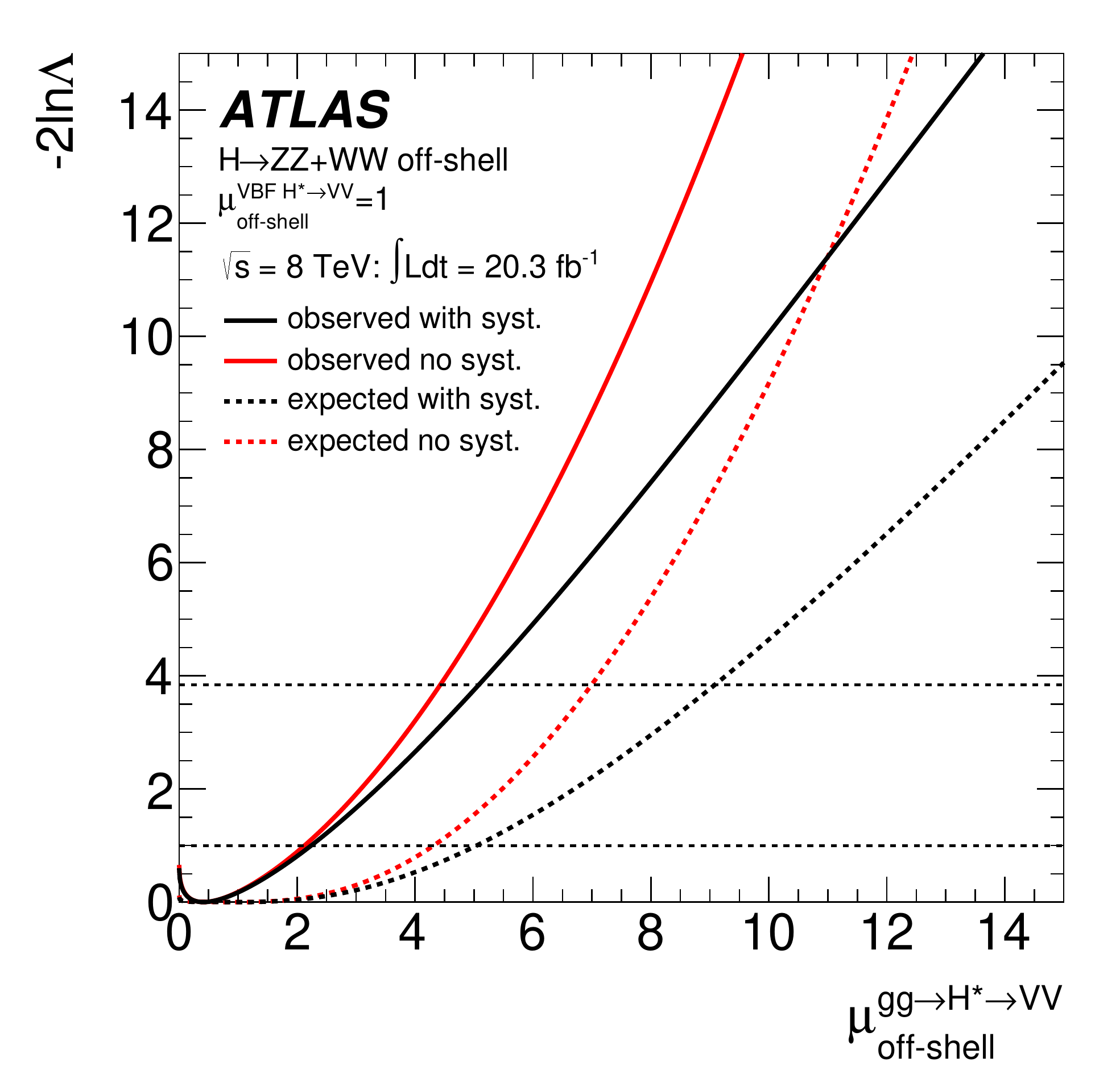}\label{fig:scanRbVBFfixed}}
\caption{Scan of the negative log-likelihood, $-2\ln\Lambda$, as a function of
\muoffshell, for the combined $ZZ$ and $WW$ analyses. 
(a) Common signal strength \muoffshell\ applied to both the $gg \to H^*$ and VBF processes.
The ratio of the $gg \to H^*$ and VBF processes is assumed to be as in the SM.
(b) Signal strength $\muoffshell^{gg \to H^* \to VV}$ for the $gg \to H^* \to VV$ process. 
The production rate for the VBF off-shell process is fixed to the SM prediction.
The black solid (dashed) line represents the observed (expected) value including all systematic uncertainties,
while the red solid (dashed) line is for the observed (expected) value without systematic uncertainties.
A relative $gg\to VV$ background K-factor of \Kratio=1 is assumed in these figures.
}
\label{fig:muoffshell_combined_scan}
\end{center}
\end{figure}

The scan of the negative log-likelihood, $-2\ln \Lambda$, as a function of
\muoffshell\ for data and the expected curve for an SM Higgs boson for the two cases above 
are shown in Fig.~\ref{fig:muoffshell_combined_scan}.

\begin{figure}[!htbp]
\begin{center}
\subfigure[]{\includegraphics[width=0.48\figwidth]{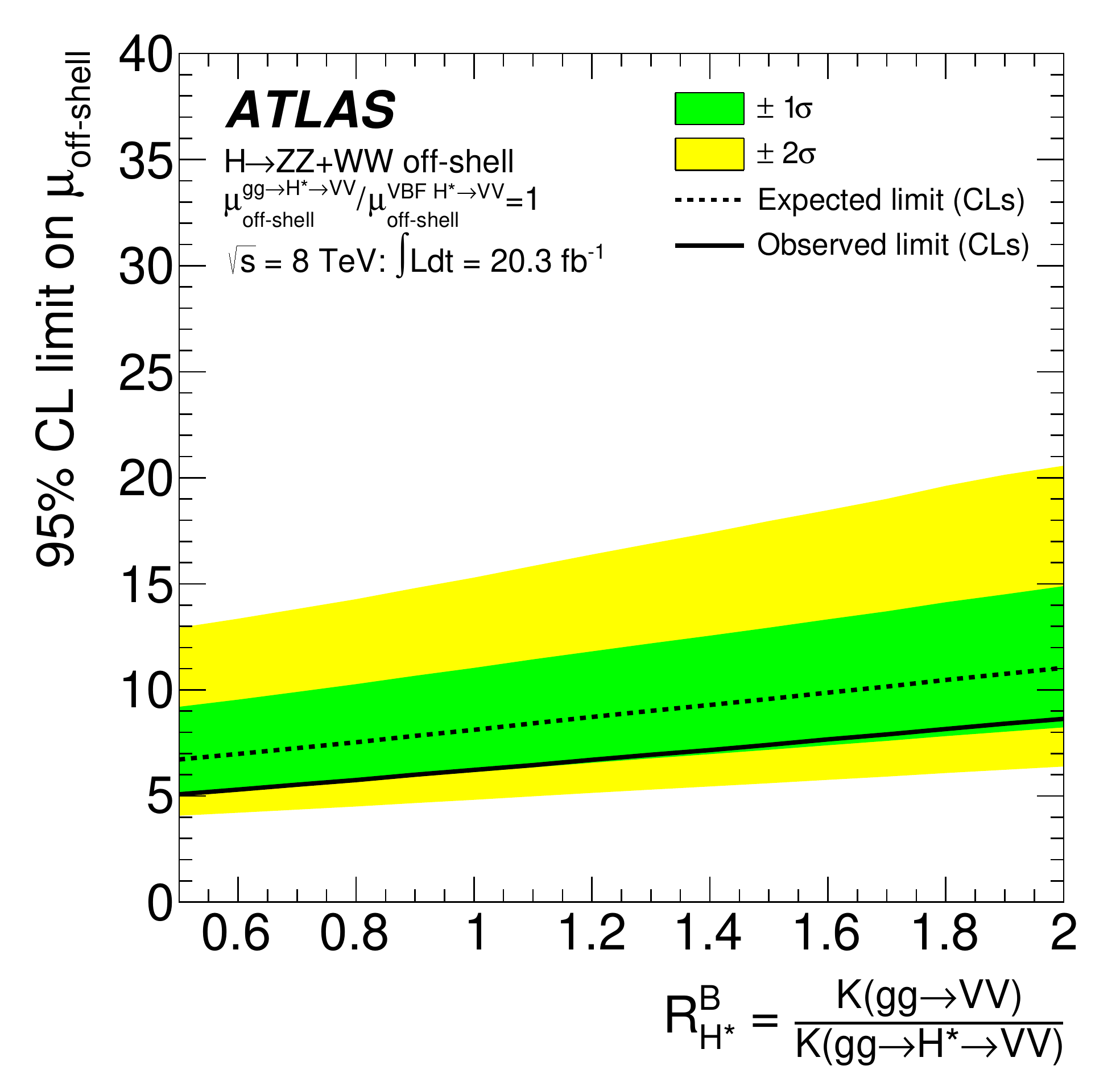}\label{fig:clsRbVBFfree}}
\subfigure[]{\includegraphics[width=0.48\figwidth]{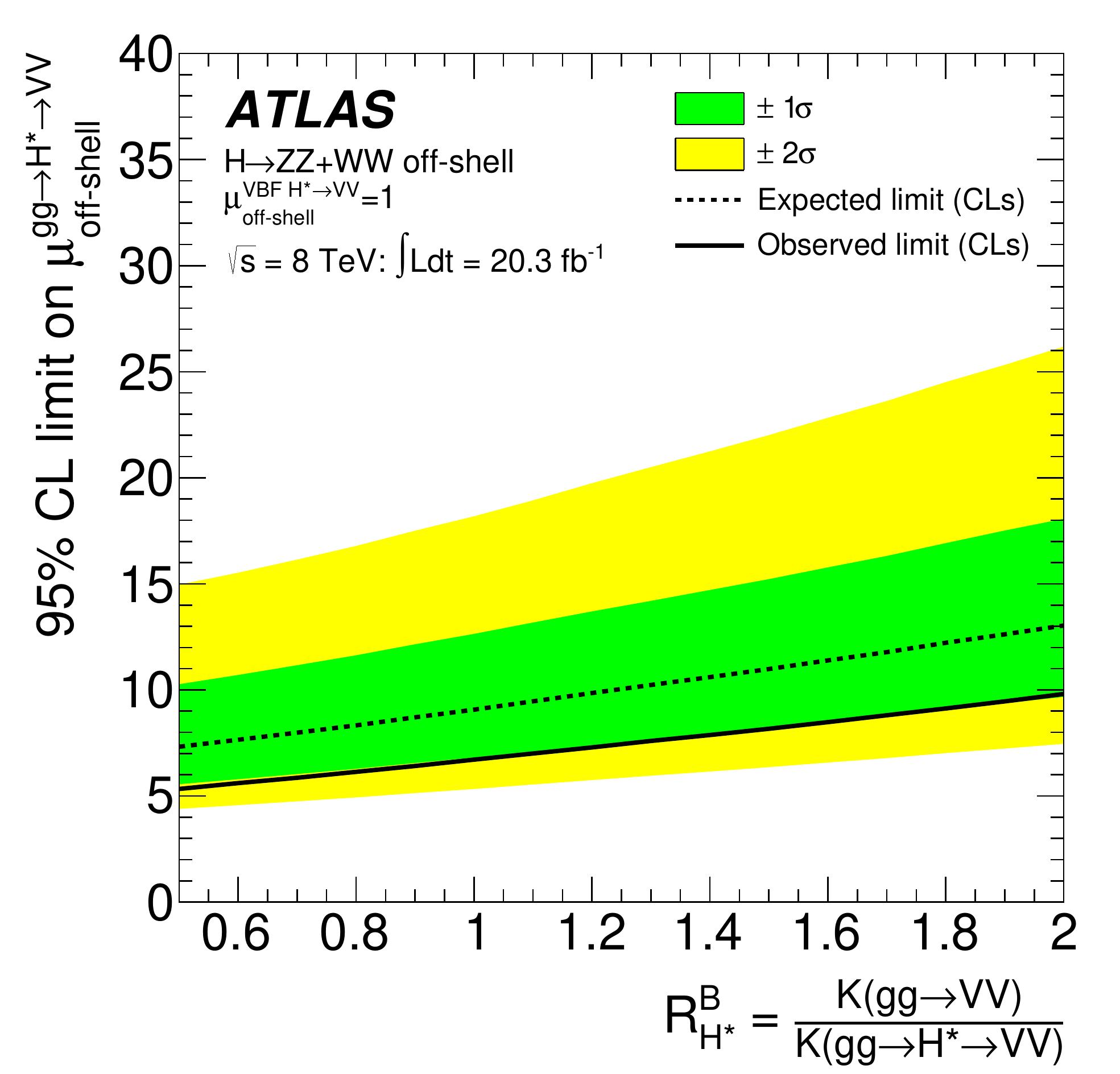}\label{fig:clsRbVBFfixed}}
\caption{The observed and expected combined 95\% CL upper limit on \muoffshell\ as a function of \Kratio\ 
for the combined $ZZ$ and $WW$ analyses. 
The upper limits are calculated using the \CLs\ method, with the SM as the alternative hypothesis. 
(a) Limit on the common signal strength \muoffshell\ applied to both the $gg \to H^*$ and VBF processes.
The ratio of the $gg \to H^*$ and VBF processes is assumed to be as in the SM.
(b) Limit on the signal strength $\muoffshell^{gg \to H^* \to VV}$ for the $gg \to H^* \to VV$ process. 
The production rate for the VBF off-shell process is fixed to the SM prediction.
The green (yellow) bands represent the 68\% (95\%) confidence intervals for the \CLs\ expected limit.
}
\label{fig:muoffshell_combined_cls}
\end{center}
\end{figure}

%%%%%%%%%%%%%%%%%%%%%%%%5
\begin{table}[!htbp]
 \centering
 \begin{tabular}{|r| r @{\hspace{4ex}}r @{\hspace{4ex}}r | r @{\hspace{4ex}}r @{\hspace{4ex}}r | l |}
  \hline
 & \multicolumn{3}{c|}{Observed} & \multicolumn{3}{c|}{Median expected} & Assumption\\
\Kratio & 0.5 & {\bf 1.0} & 2.0  & 0.5 & {\bf 1.0} & 2.0 & \\ [0.5ex]
\hline
\rule[-1.1ex]{0pt}{3.6ex}
$\mu_\text{\text{off-shell}}$ & 5.1 & {\bf 6.2} & 8.6 & 6.7 & {\bf 8.1} & 11.0 & $\mu_{\text{off-shell}}^{gg \to H^*} / \mu_{\text{off-shell}}^{VBF}$=1\\
\rule[-1.1ex]{0pt}{3.6ex}
$\mu_\text{\text{off-shell}}^{gg \to H^* \to VV}$ & 5.3 & {\bf 6.7} & 9.8 & 7.3 & {\bf 9.1} & 13.0 & $\mu_{\text{off-shell}}^{\text{VBF }H^* \to VV}$=1\\
\hline
 \end{tabular}
  \caption{The observed and expected 95\% CL upper limits on $\mu_\text{off-shell}$  and $\mu_\text{off-shell}^{gg \to H^* \to VV}$ 
within the range of $0.5<\Kratio<2$ for the combined $ZZ$ and $WW$ analyses. 
Results are shown for two hypotheses, which are defined in the Assumption column.
The bold numbers correspond to the limit assuming $\Kratio=1$.
The upper limits are evaluated using the \CLs\ method, 
with the alternative hypothesis $\mu_\text{off-shell}=1$.}
\label{tab:limitOFFshell_ZZWW}
\end{table}
The limits on $\mu_{\text{off-shell}}$ and $\mu_{\text{off-shell}}^{gg\to H^*}$ are computed with the \CLs\ method, 
assuming for the alternative hypothesis that all the off-shell rates are at their SM predictions.
They are derived as a function of the $gg \to VV$ background K-factor ratio \Kratio. These results are 
reported in Table~\ref{tab:limitOFFshell_ZZWW} and shown in Fig.~\ref{fig:muoffshell_combined_cls}, assuming 
either one common scale factor for both the $gg \to H^*$ and VBF processes 
or using a scale factor for the $gg \to H^*$ process and fixing the VBF production to the SM prediction. 

%%%%%%%%%%%%%%%%
%%%%%%%%%%%%%%%%

%%%%%%%%%%%%%%%%
%%%%%%%%%%%%%%%%

The impact of the various systematic uncertainties on the combined expected limit in the off-shell fit are listed in 
Table \ref{tab:nuisance_offShell_conf2} when fixing the ratio of the signal strength in $gg \to H^*$ and VBF to 
the SM prediction. The values in this table were derived by fixing all the nuisance parameters associated with the systematic uncertainties 
to the values derived from the SM-conditional fit to the data, with the exception of the one under study.

%%%%%%%%%%%%%%%%%%%
\begin{table}[!ht]
\begin{center}
\begin{tabular}{|l|c|}
\hline
Systematic uncertainty &  95\% CL lim. $(CL_s)$ on $\mu_{\mathrm{off-shell}}$ \\
\hline
Interference $gg \to (H^* \to) VV$ & 7.2 \\
QCD scale \KHmVV\ (correlated component) & 7.1\\
PDF $q\bar{q} \to VV$ and $gg \to (H^* \to) VV$ & 6.7\\
QCD scale $q\bar{q} \to VV$ & 6.7\\
Luminosity & 6.6\\
Drell--Yan background & 6.6\\
QCD scale \KHggmVV\ (uncorrelated component) & 6.5\\
Remaining systematic uncertainties & 6.5\\
\hline
$\textbf{All systematic uncertainties}$  & 8.1 \\
\hline
$\textbf{No systematic uncertainties}$  & 6.5\\
\hline
\end{tabular}
\caption{The expected 95\% CL upper limit on $\mu_\text{off-shell}$ for the combined $ZZ$ and $WW$ analyses, 
with a ranked listing of each systematic uncertainty individually, comparing with 
no systematic uncertainty or all systematic uncertainties.  
The upper limits are evaluated using the \CLs\ method, assuming \Kratio=1. 
The ratio of the $gg \to H^*$ and VBF processes is assumed to be as expected in the SM.}
\label{tab:nuisance_offShell_conf2}
\end{center}
\end{table}

%%%%%%%%%%%%%%%%%%

\subsection{Combination of the off-shell and on-shell $ZZ$ and $WW$ analyses}

In this section, the off-shell results reported above are combined with the on-shell $H \to ZZ^{*} \to 4 \ell$ \cite{Aad:2014eva} and $H \to WW^{*} \to \ell \nu \ell \nu$ \cite{ATLAS:2014aga}
analyses based on the 8~\TeV\ data taken in 2012. In these analyses a Higgs boson mass value of $125.36$~\GeV~\cite{Aad:2014aba} is assumed. 
For the on-shell $ZZ$ and $WW$ combination the main common sources of theoretical and experimental systematic uncertainties are treated as correlated~\cite{Aad:2013wqa}.

The uncertainties from the impact of higher-order QCD corrections on the $gg \to H^{(*)}$ and $qq \to VV$ processes
are considered correlated between the on-shell and off-shell measurements. 
The PDF uncertainties are treated as uncorrelated between on-shell and off-shell analyses. 
The correlations between the PDF uncertainties for the on-shell and off-shell analyses 
are expected to be small with the exception of the ones for the $q\bar{q}\to VV$ process, 
which have negligible impact on the on-shell results.
  
In addition to the main theoretical uncertainties, 
the common experimental systematic uncertainties are treated as correlated.

The results reported in the following are based on two different assumptions:
\begin{itemize}
\item Determination of \Widthratio\ when profiling the coupling scale factors $\kappa_{g}$ and $\kappa_{V}$ associated with the on- and off-shell $gg \to H^{(*)}$ and VBF production and the $H^{(*)} \to VV$ decay, 
assuming $\kappa_{g}=\kappa_{g,\text{on-shell}}=\kappa_{g,\text{off-shell}}$ and $\kappa_{V}=\kappa_{V,\text{on-shell}}=\kappa_{V,\text{off-shell}}$.
\item Determination of $R_{gg}=\kappa^2_{g,\text{off-shell}}/\kappa^2_{g,\text{on-shell}}$ when profiling the coupling scale factor $\kappa_{V}=\kappa_{V,\text{on-shell}}=\kappa_{V,\text{off-shell}}$ associated with the VBF production and the $H^{(*)} \to VV$ decay. The ratio \Widthratio=1 is fixed to the SM prediction. 
The parameter $R_{gg}$ is sensitive to possible modifications of the gluon couplings in the high-mass range with respect to the on-shell value.
\end{itemize}

\begin{figure}[!htbp]
\begin{center}
\subfigure[]{\includegraphics[width=0.47\figwidth]{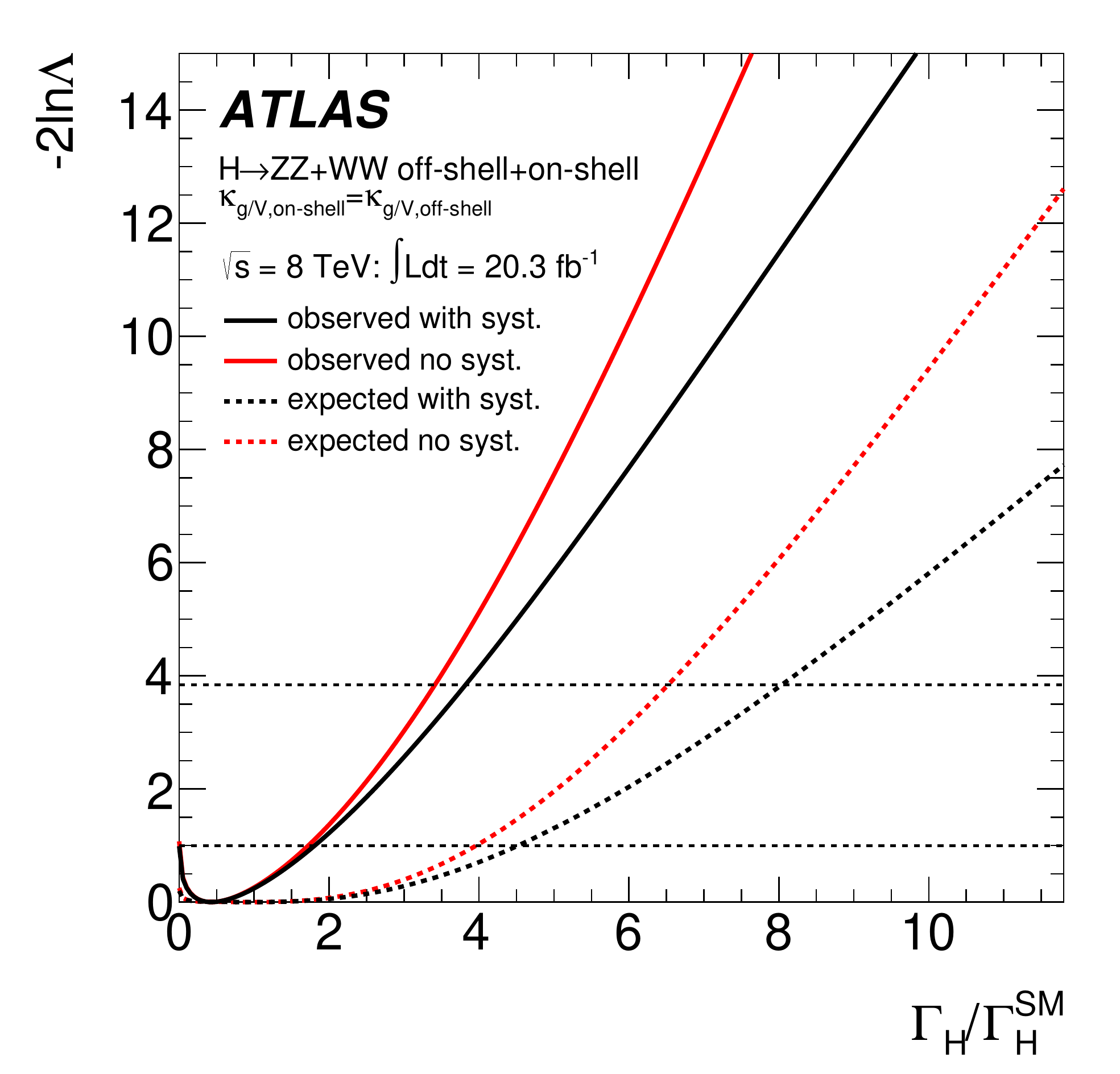}}
\subfigure[]{\includegraphics[width=0.47\figwidth]{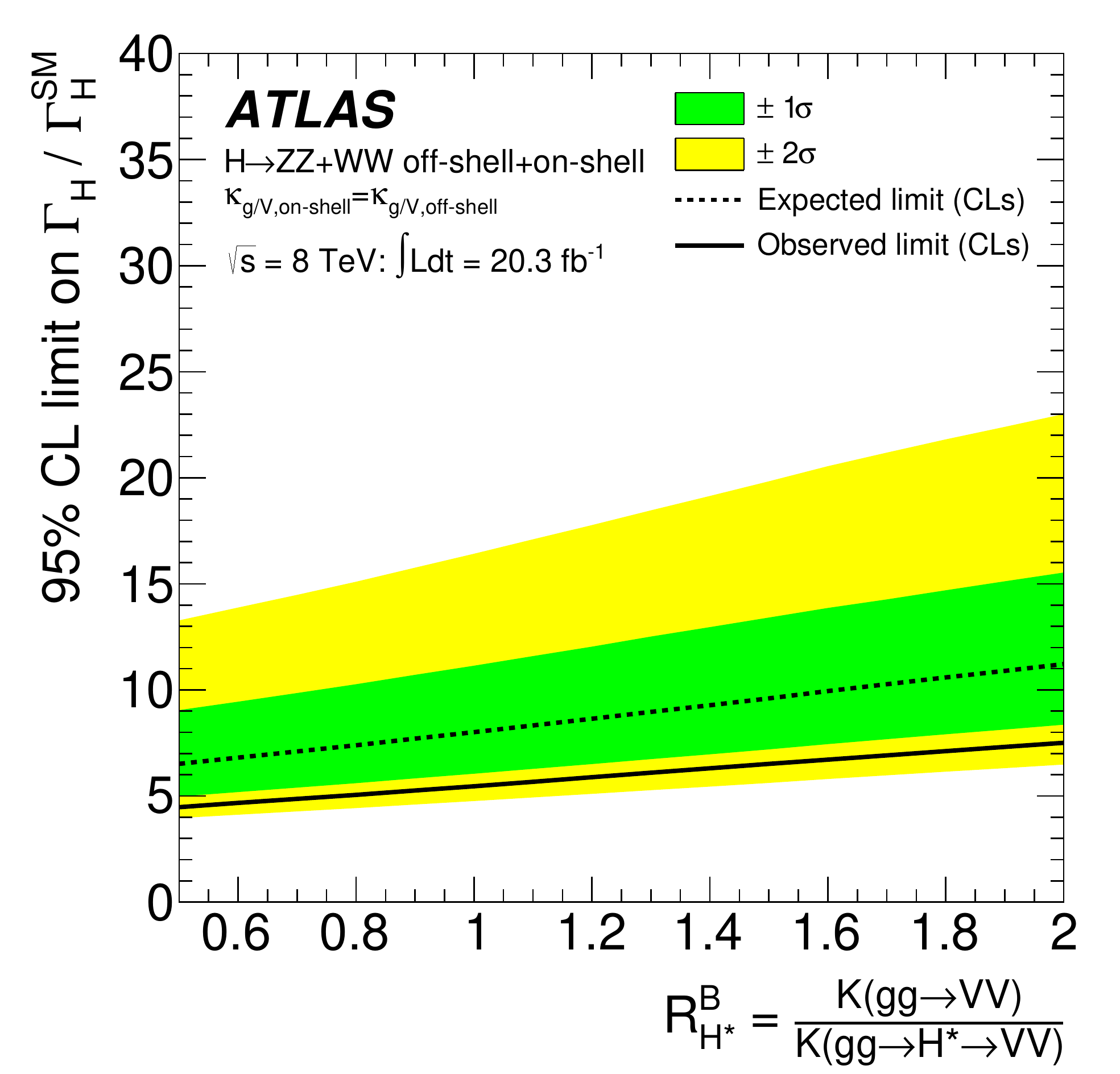}}
\caption{(a) Scan of the negative log-likelihood as a function of \Widthratio\ when profiling the coupling scale 
factors $\kappa_{g}$ and $\kappa_{V}$ associated with the on- and off-shell $gg \to H^{(*)}$ and VBF production and the $H^{(*)} \to VV$ decay. 
The black solid (dashed) line represents the observed (expected) value including all systematic uncertainties,
while the red solid (dashed) line is for the observed (expected) value without systematic uncertainties.
(b) Observed and expected combined 95\% CL upper limit on \Widthratio\ as a function of \Kratio\ under the same assumption as (a). 
The upper limits are calculated from the \CLs\ method, with the SM values as the alternative hypothesis.
The green (yellow) bands represent the 68\% (95\%) confidence intervals for the \CLs\ expected limit.
}
\label{fig:combALL_conf3}
\end{center}
\end{figure}

\begin{figure}[!htbp]
\begin{center}
\subfigure[]{\includegraphics[width=0.47\figwidth]{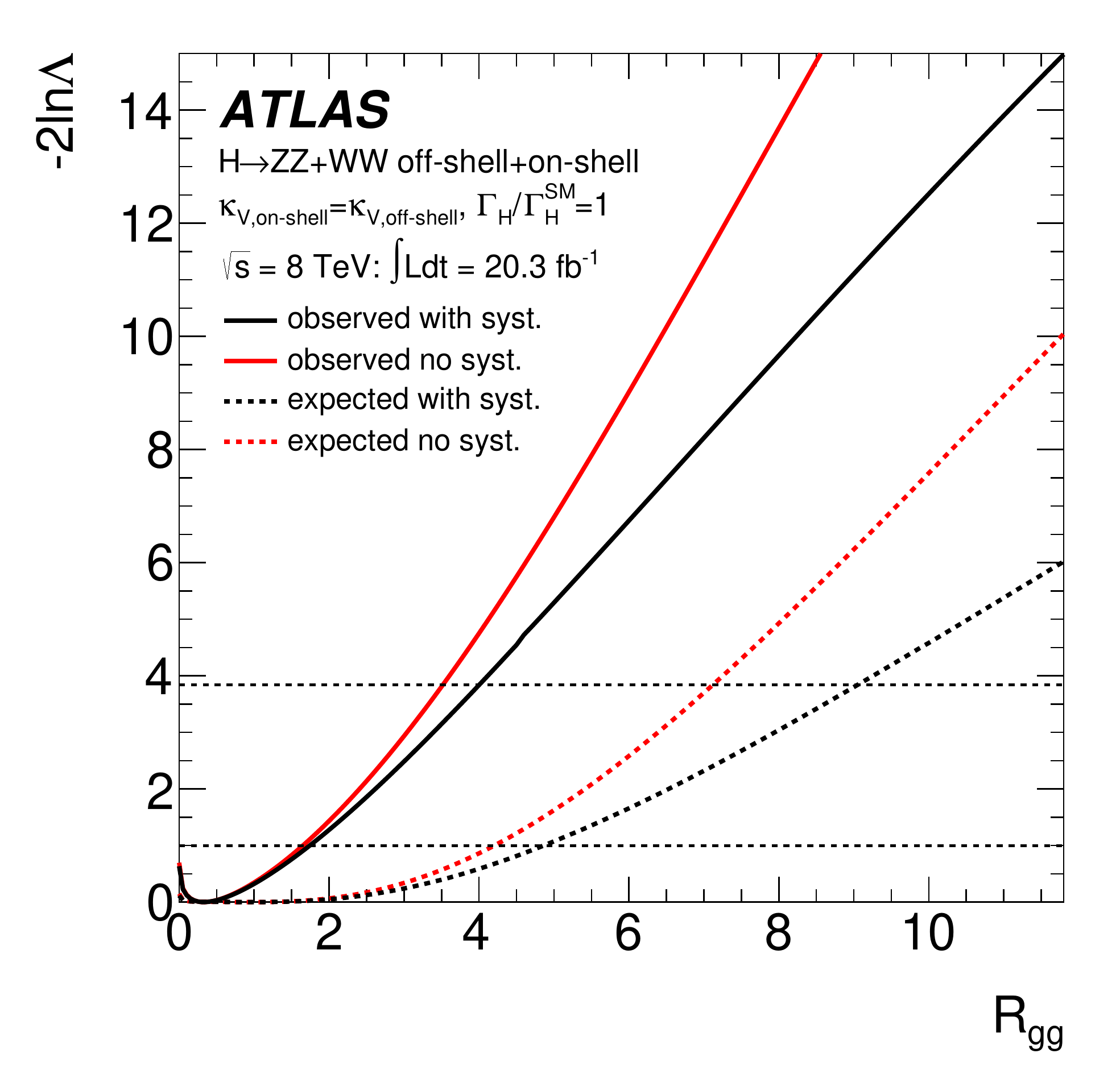}}
\subfigure[]{\includegraphics[width=0.47\figwidth]{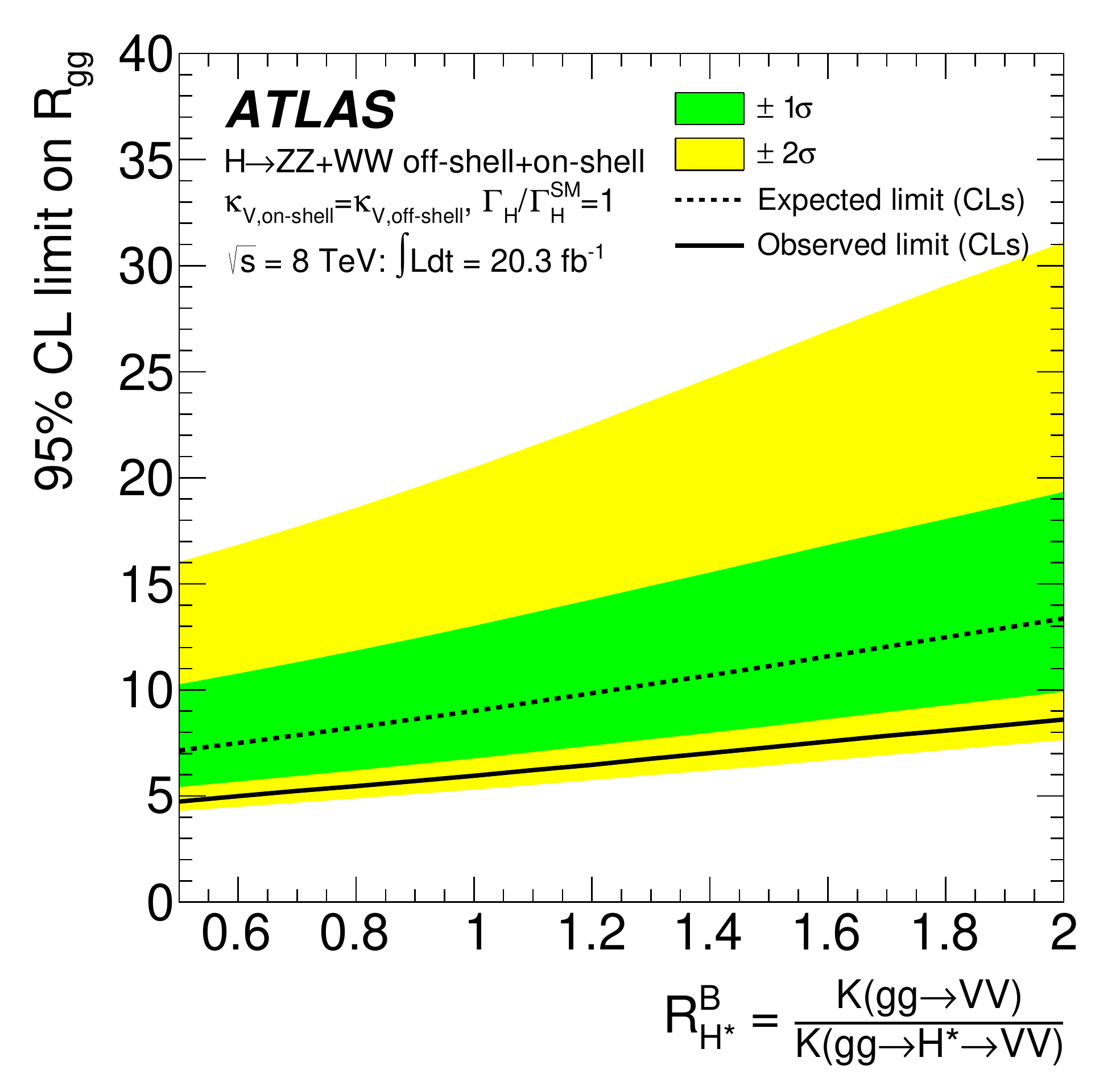}}
\caption{(a) Scan of $R_{gg}=\kappa^2_{g,\text{off-shell}}/\kappa^2_{g,\text{on-shell}}$ 
when profiling the coupling scale factor $\kappa_{V}$ associated with the on- and off-shell  
VBF production and the $H^{(*)} \to VV$ decay. The ratio \Widthratio\ is set to 1.0. 
The black solid (dashed) line represents the observed (expected) value including all systematic uncertainties,
while the red solid (dashed) line is for the observed (expected) value without systematic uncertainties.
(b) Observed and expected combined 95\% CL upper limit on $R_{gg}$ as a function of \Kratio\ under the same assumption as (a). 
The upper limits are calculated from the \CLs\ method, with the SM values as the alternative hypothesis.
The green (yellow) bands represent the 68\% (95\%) confidence intervals for the \CLs\ expected limit.}
\label{fig:combALL_conf4}
\end{center}
\end{figure}

The negative log-likelihood scans for the above-defined fitting configurations as well as the combined upper 
limit at 95\% CL on \Widthratio\ and $R_{gg}$ are illustrated in Figs.~\ref{fig:combALL_conf3} and \ref{fig:combALL_conf4} and 
the corresponding limits are listed in Table~\ref{tab:obs_on_off_combined_limit}.
The limits are all computed with the \CLs\ method, taking the SM values as the alternative hypothesis. 

The limit on \Widthratio\ can be translated into a limit on the total width of the Higgs boson under the assumptions reported above, 
out of which the most important is that the relevant Higgs boson coupling scale factors are independent of the energy scale of the Higgs boson production. 
Assuming a value of \Kratio=1, this translates into an observed (expected) 95\% CL upper limit on the Higgs boson total width
of 22.7 (33.0) MeV.\footnote{The value of the SM Higgs boson width of 4.12 MeV
at a mass of 125.4 GeV~\cite{Heinemeyer:2013tqa} is used to convert the limit  \Widthratio\ into the total width limit.} 

%%%%%%%%%%%%%%%%
%%%%%%%%%%%%%%%%

%%%%%%%%%%%%%%%%
%%%%%%%%%%%%%%%%

%%%%%%%%%%%%%%%%%%%%%%%%%%
\begin{table}[!ht]
\begin{center}
 \begin{tabular}{|r| r @{\hspace{4ex}}r @{\hspace{4ex}}r | r @{\hspace{4ex}}r @{\hspace{4ex}}r | l |}
\hline
 & \multicolumn{3}{c|}{Observed} & \multicolumn{3}{c|}{Median expected} & Assumption \\
\Kratio & 0.5 & {\bf 1.0} & 2.0  & 0.5 & {\bf 1.0} & 2.0 &  \\ [0.5ex]
\hline
\rule[-1.1ex]{0pt}{3.6ex}
\Widthratio\ & 4.5   & {\bf 5.5}   & 7.5  & 6.5  & {\bf 8.0}  & 11.2 & $\kappa_{i,\text{on-shell}}=\kappa_{i,\text{off-shell}}$ \\
$R_{gg}=\kappa^2_{g,\text{off-shell}}/\kappa^2_{g,\text{on-shell}}$ & 4.7   & {\bf 6.0}   & 8.6 & 7.1  &  {\bf 9.0}  & 13.4 & $\kappa_{V,\text{on-shell}}=\kappa_{V,\text{off-shell}}$, \Widthratio=1 \\
\hline
\end{tabular}
\caption{Observed and expected 95\% CL upper limits on \Widthratio\ and $R_{gg}$ for the combined on- and off-shell $ZZ$ and $WW$ analyses.
Results are shown for two hypotheses, which are defined in the Assumption column.
\Kratio\ is within the range 0.5$<$ \Kratio $<$ 2.}
\label{tab:obs_on_off_combined_limit}
\end{center}
\end{table}
%%%%%%%%%%%%%%%%%%%%%%%%%%

%-------------------------------------------------------------------------------
\section{Conclusion}
\label{sec:conclusion}
The measurement of the $ZZ$ and $WW$ final states in the mass range above the $2m_Z$ and $2m_W$ thresholds
provides a unique opportunity to measure the off-shell coupling strengths of the observed Higgs boson.
In this paper constraints on the off-shell Higgs boson signal strengths in the $ZZ \to 4\ell$, $ZZ\to2\ell2\nu$ and $\ww$
final states and their combination are presented.
The result is based on $pp$ collision data collected by the ATLAS experiment at the LHC,
corresponding to an integrated luminosity of $20.3$~fb$^\text{-1}$ at a collision energy of $\sqrt{s}=8~\TeV$.

Using the  \CLs\ method, the observed 95\% confidence level (CL) upper limit on the off-shell signal strength is in the range
5.1--8.6, with an expected range of 6.7--11.0. In each case the range is determined by varying the unknown $gg\to ZZ$ and 
$gg\to WW$ background K-factor from higher-order QCD corrections between half and twice the value of the known signal K-factor. 

Assuming the relevant Higgs boson couplings are independent of the energy scale of the Higgs production,
a combination with the on-shell measurements of $ZZ$ and $WW$ in the same dataset yields  
an observed (expected) 95\% CL upper limit on \Widthratio\ 
in the range 4.5--7.5 (6.5--11.2) 
under the same variations of the background K-factor. 
Assuming the value of \Kratio=1 and under the assumptions reported above, 
this translates into an observed (expected) 95\% CL 
upper limit on the Higgs boson total width of 22.7 (33.0) MeV. 

Assuming that the total width of the Higgs boson is as expected in 
the SM, the same combination can be interpreted as a limit on the ratio of the off-shell 
to the on-shell couplings to gluons $R_{gg} = \kappa^2_{g,\text{off-shell}}/\kappa^2_{g,\text{on-shell}}$. 
An observed (expected) 95\% CL upper 
limit on $R_{gg}$ in the range 4.7--8.6 (7.1--13.4)
under the same variations of the background K-factor is found.

%-------------------------------------------------------------------------------
\section*{Acknowledgments}
\label{sec:acknowledgements}
We are very thankful to M.~Bonvini, J.~Campbell, S.~Forte, F.~Krauss, K.~Melnikov, G.~Passarino, and M.~Spannowsky
for their essential input in the estimation of uncertainties in the theoretical predictions for the signal and background 
processes and their interference.

% Acknowledgements for papers with collision data
% Version 23-Mar-2015

% Standard acknowledgements start here
%----------------------------------------------
We thank CERN for the very successful operation of the LHC, as well as the
support staff from our institutions without whom ATLAS could not be
operated efficiently.

We acknowledge the support of ANPCyT, Argentina; YerPhI, Armenia; ARC,
Australia; BMWFW and FWF, Austria; ANAS, Azerbaijan; SSTC, Belarus; CNPq and FAPESP,
Brazil; NSERC, NRC and CFI, Canada; CERN; CONICYT, Chile; CAS, MOST and NSFC,
China; COLCIENCIAS, Colombia; MSMT CR, MPO CR and VSC CR, Czech Republic;
DNRF, DNSRC and Lundbeck Foundation, Denmark; EPLANET, ERC and NSRF, European Union;
IN2P3-CNRS, CEA-DSM/IRFU, France; GNSF, Georgia; BMBF, DFG, HGF, MPG and AvH
Foundation, Germany; GSRT and NSRF, Greece; RGC, Hong Kong SAR, China; ISF, MINERVA, GIF, I-CORE and Benoziyo Center, Israel; INFN, Italy; MEXT and JSPS, Japan; CNRST, Morocco; FOM and NWO, Netherlands; BRF and RCN, Norway; MNiSW and NCN, Poland; GRICES and FCT, Portugal; MNE/IFA, Romania; MES of Russia and NRC KI, Russian Federation; JINR; MSTD,
Serbia; MSSR, Slovakia; ARRS and MIZ\v{S}, Slovenia; DST/NRF, South Africa;
MINECO, Spain; SRC and Wallenberg Foundation, Sweden; SER, SNSF and Cantons of
Bern and Geneva, Switzerland; NSC, Taiwan; TAEK, Turkey; STFC, the Royal
Society and Leverhulme Trust, United Kingdom; DOE and NSF, United States of
America.

The crucial computing support from all WLCG partners is acknowledged
gratefully, in particular from CERN and the ATLAS Tier-1 facilities at
TRIUMF (Canada), NDGF (Denmark, Norway, Sweden), CC-IN2P3 (France),
KIT/GridKA (Germany), INFN-CNAF (Italy), NL-T1 (Netherlands), PIC (Spain),
ASGC (Taiwan), RAL (UK) and BNL (USA) and in the Tier-2 facilities
worldwide.
%----------------------------------------------

\appendix
\section{Monte Carlo and PDF scaling to arbitrary $\mu_\text{off-shell}$}
\def\muoffshell{\ensuremath{\mu_\text{off-shell}}}

The known dependence of the off-shell Higgs boson signal process, the background process and the interference term on the off-shell signal strength
$\muoffshell$ can be used to construct MC samples for arbitrary values of $\muoffshell$ from three basic samples generated at different fixed
values of $\muoffshell$.

\subsection{Dependence of the $gg\to(H^* \to)VV$ off-shell cross-sections on the signal strength}

\label{Sec:App:ggHVVscaling}
An event sample $\sigma_{gg\rightarrow (H^* \rightarrow) VV}(\mu_\text{off-shell})$ for the 
$gg \to (H^* \to) VV$ process with an arbitrary value of the 
off-shell Higgs boson signal strength $\mu_\text{off-shell}$ can be constructed from the MC sample 
for the SM Higgs boson signal $gg \to H^* \to VV$ ($\sigma_{gg\rightarrow H^*\rightarrow VV}^\text{SM}$), 
the $gg \to VV$ continuum background MC sample ($\sigma_{gg\rightarrow VV,\,\text{cont}}$) and 
a full SM Higgs boson signal plus background $gg \to (H^*\to) VV$ MC sample 
($\sigma_{gg\rightarrow (H^* \rightarrow)VV}^\text{SM}$) using the following weighting function:
%%%%%%%%%%%%%%%%%%%%
\begin{eqnarray}
\label{eqn:ggHZZscaling_on_S_B_I}
\sigma_{gg\rightarrow (H^* \rightarrow) VV}(\mu_\text{off-shell},m_{VV})& = & 
\KHmVV \cdot \mu_\text{off-shell} \cdot \sigma_{gg\rightarrow H^*\rightarrow VV}^\text{SM}(m_{VV}) \\\nonumber
& + & \sqrt{ \KHggmVV \cdot \KBmVV \cdot \mu_\text{off-shell} } \cdot \sigma_{gg\rightarrow VV,\, \text{Interference}}^{\text{SM}}(m_{VV}) \\\nonumber
& + & \KBmVV \cdot \sigma_{gg\rightarrow VV,\, \text{cont}}(m_{VV}) \,, \\
\label{eqn:def_interference}
\sigma_{gg\rightarrow VV,\ \text{Interference}}^{\text{SM}}(m_{VV}) &=& \sigma_{gg\rightarrow (H^* \rightarrow)VV}^\text{SM}(m_{VV}) 
                                                 -\sigma_{gg\rightarrow H^*\rightarrow VV}^\text{SM}(m_{VV})  
                                                 -\sigma_{gg\rightarrow VV,\,\text{cont}}(m_{VV}) \,,
\end{eqnarray}
where the K-factors are calculated inclusively without any selections. 

As a direct simulation of an interference MC sample is not possible, Eq.~(\ref{eqn:def_interference}) and \Kratio\ are used to obtain:
\begin{eqnarray}
\label{eqn:ggHZZscaling_on_S_SBI_B}
\sigma_{gg\rightarrow (H^* \rightarrow) VV}(\mu_\text{off-shell},m_{VV}) & = & \left( \KHmVV \cdot \mu_\text{off-shell} - \KHggmVV \cdot \sqrt{\Kratio \cdot \mu_\text{off-shell}}\right) \cdot \sigma_{gg\rightarrow H^*\rightarrow VV}^\text{SM}(m_{VV}) \nonumber \\ 
& + & \KHggmVV \cdot \sqrt{\Kratio\cdot \mu_\text{off-shell}} \cdot \sigma_{gg\rightarrow (H^* \rightarrow) VV}^\text{SM}(m_{VV}) \\ \nonumber
& + & \KHggmVV \cdot \left(\Kratio-\sqrt{\Kratio\cdot \mu_\text{off-shell}}\right) \cdot \sigma_{gg\rightarrow VV,\,\text{cont}}(m_{VV}) .
\end{eqnarray}
%%%%%%%%%%%%%%%%%%%%

\subsection{Dependence of the $jjVV$ off-shell signal and background interference on the signal strength}
\label{Sec:App:VBFHVVscaling}
An MC event sample for the EW $pp\to (H^* + 2j \rightarrow) VV + 2j$ process with an arbitrary value of the 
off-shell Higgs boson signal strength $\mu_\text{off-shell}$ can be constructed from a pure $pp\to VV + 2j$ continuum background 
MC sample, a full SM Higgs boson signal plus background $pp\to (H^* + 2j \rightarrow) VV + 2j$ MC sample and 
a third Higgs boson signal plus background $pp\to (H^* + 2j \rightarrow) VV + 2j$ MC sample with 
$\mu_\text{off-shell}=\kappa^4_{V}=\Widthratio=10$. 
Using $\Widthratio=10$ for the last sample ensures that the on-shell $VH$ events are generated with SM-like signal strength.

The following weighting function is used:
%%%%%%%%%%%%%%%%%%%%
\begin{eqnarray}
\label{eqn:VBFscaling_on_S_B_I}
\sigma_{pp\rightarrow (H^* + 2j \rightarrow) VV + 2j}(\mu_\text{off-shell})& = & 
\mu_\text{off-shell} \cdot \sigma_{pp\rightarrow H^* + 2j \rightarrow VV + 2j}^\text{SM} \\\nonumber
& + & \sqrt{ \mu_\text{off-shell} } \cdot \sigma_{pp\rightarrow VV + 2j,\ \text{Interference}} \\\nonumber
& + & \sigma_{pp\rightarrow VV + 2j,\,\text{cont}} \,, 
\end{eqnarray}
where the signal and interference samples are implicitly defined through the SM \linebreak
$pp\to (H^* + 2j \rightarrow) VV + 2j$ MC sample
\begin{eqnarray}
\label{eqn:VBFdef_interference}
\sigma_{pp\rightarrow (H^* + 2j \rightarrow) VV + 2j}^\text{SM} & = & 
\sigma_{pp\rightarrow H^* + 2j \rightarrow VV + 2j}^\text{SM} + 
\sigma_{pp\rightarrow VV + 2j,\ \text{Interference}} + 
\sigma_{pp\rightarrow VV + 2j,\,\text{cont}} 
\end{eqnarray}
and a $\muoffshell=10$ MC sample:
\begin{eqnarray}
\sigma_{pp\rightarrow (H^* + 2j \rightarrow) VV + 2j}^{\kappa^4_{V}=10} & = & 
10 \cdot \sigma_{pp\rightarrow H^* + 2j \rightarrow VV + 2j}^\text{SM} +
\sqrt{ 10 } \cdot \sigma_{pp\rightarrow VV + 2j,\ \text{Interference}} +
\sigma_{pp\rightarrow VV + 2j,\,\text{cont}} \, .
\end{eqnarray}
Solving for the generated MC samples yields:
\begin{eqnarray}
\label{eq:pdf_VBFzz}
\sigma_{pp\rightarrow (H^* + 2j \rightarrow) VV + 2j}(\mu_\text{off-shell})& = & 
      \frac{ \mu_\text{off-shell} - \sqrt{\mu_\text{off-shell}} }{10-\sqrt{10}} 
      \sigma_{pp\rightarrow (H^* + 2j \rightarrow) VV + 2j}^{\kappa^4_{V}=10}  \\ \nonumber
& + & \frac{10\sqrt{\mu_\text{off-shell}} - \sqrt{10}\mu_\text{off-shell} }{10-\sqrt{10}} 
      \sigma_{pp\rightarrow (H^* + 2j \rightarrow) VV + 2j}^\text{SM} \\\nonumber
& + & \frac{(\sqrt{\mu_\text{off-shell}}-1)\cdot(\sqrt{\mu_\text{off-shell}}-\sqrt{10})}{\sqrt{10}} 
      \sigma_{pp\rightarrow VV + 2j,\,\text{cont}}. 
\end{eqnarray}
%%%%%%%%%%%%%%%%%%%%

\bibliographystyle{atlasBibStyleWithTitle}
\bibliography{width}

\onecolumn
\clearpage
% ATLAS Collaboration author list
% Data extracted on 27-Mar-2015 for paper reference HIGG-2014-10

\begin{flushleft}
{\Large The ATLAS Collaboration}

\bigskip

G.~Aad$^{\rm 85}$,
B.~Abbott$^{\rm 113}$,
J.~Abdallah$^{\rm 152}$,
O.~Abdinov$^{\rm 11}$,
R.~Aben$^{\rm 107}$,
M.~Abolins$^{\rm 90}$,
O.S.~AbouZeid$^{\rm 159}$,
H.~Abramowicz$^{\rm 154}$,
H.~Abreu$^{\rm 153}$,
R.~Abreu$^{\rm 30}$
% ...
\bigskip
\\
% ...
$^{11}$ Institute of Physics, Azerbaijan Academy of Sciences, Baku, Azerbaijan\\
% ...
$^{30}$ CERN, Geneva, Switzerland\\
% ...
$^{85}$ CPPM, Aix-Marseille Universit{\'e} and CNRS/IN2P3, Marseille, France\\
% ...
$^{90}$ Department of Physics and Astronomy, Michigan State University, East Lansing MI, United States of America\\
% ...
$^{107}$ Nikhef National Institute for Subatomic Physics and University of Amsterdam, Amsterdam, Netherlands\\
% ...
$^{113}$ Homer L. Dodge Department of Physics and Astronomy, University of Oklahoma, Norman OK, United States of America\\
% ...
$^{152}$ Institute of Physics, Academia Sinica, Taipei, Taiwan\\
$^{153}$ Department of Physics, Technion: Israel Institute of Technology, Haifa, Israel\\
$^{154}$ Raymond and Beverly Sackler School of Physics and Astronomy, Tel Aviv University, Tel Aviv, Israel\\
% ...
$^{159}$ Department of Physics, University of Toronto, Toronto ON, Canada\\
% ...
\end{flushleft}

% Created with xml2latex.py

\end{document}